\documentclass[preprint,aps,nofootinbib]{revtex4-2}
\pdfoutput=1
\usepackage{graphicx}
\usepackage{epstopdf}    
\usepackage{dcolumn}
\usepackage{amsfonts}
\usepackage{amsmath}
\usepackage{amssymb}
\usepackage{bm}
\usepackage{color}
\usepackage{comment}
\usepackage[caption=false]{subfig}
\usepackage{lipsum} 
\usepackage{epsf}
\usepackage{float}
\usepackage{enumitem}
\usepackage[utf8]{inputenc}

\newcolumntype{L}{>{\centering\arraybackslash}m{3cm}}

\newcommand{\edc}{\end{document}}
\newcommand{\bb} {\begin{thebibliography}}
\newcommand{\eb}{\end{thebibliography}}
\newcommand{\bi}[1]{\bibitem{#1}}
\newcommand{\bc}{\begin{center}}
\newcommand{\ec}{\end{center}}

\newcommand{\be}{\begin{equation}\small}
\newcommand{\ee}{\end{equation}\normalsize}
\newcommand{\bea}{\begin{eqnarray}}
\newcommand{\eea}{\end{eqnarray}}
\newcommand{\ba}{\begin{array}{l}   }
\newcommand{\lab}[1]{\label{#1}}
\newcommand{\ea}{\end{array}}

\newcommand{\dsfrac}{\displaystyle\frac}

\newcommand{\re}[1]{(\ref{#1})}

\newcommand{\ci}{\cite}

\newcommand{{\vergul}}{  ,}

\newcommand{\veps}{\varepsilon }

\begin{document}
\draft
\title{Self-consistent theory of a homogeneous binary Bose mixture with strong repulsive
interspecies interaction}
\author{Abdulla Rakhimov$^1$, Tolibjon Abdurakhmonov$^{2,3}$, Zabardast Narzikulov$^1$ and Vyacheslav I. Yukalov$^{4,5}$}


\affiliation{
$^1$Institute of Nuclear Physics, Tashkent 100214, Uzbekistan \\
$^2$Physical-Technical Institute, Tashkent, 100084, Uzbekistan \\
$^3$University of Rostock, Rostock,  D-18059,  Germany\\
$^4$Bogoliubov Laboratory of Theoretical Physics, Joint Institute for Nuclear Research, Dubna 141980, Russia \\
$^5$Instituto de Fisica de S\~ao Carlos, Universidade de S\~ao Paulo, \\
Caixa Postale 369,  13560-970, S\~ao Carlos, S\~ao Paulo, Brazil
}

\date{\today}

\begin{abstract}
Multicomponent quantum gases are ideal platforms to study fundamental phenomena arising
from the mutual interaction between different constituents. Particularly, due to the
repulsive interactions between two species, the system may exhibit a phase separation.
We develop a mean-field-based theory for a two-component Bose mixture, which is
equivalent to the Hartree-Fock-Bogoliubov approximation, and derive analytical
expressions for the phase boundary and miscibility. The majority of existing
theories, which are valid only for weakly interacting Bose gases, predict that the phase
boundary is determined by the criterion $g_{ab}\leqslant\sqrt{g_{aa} g_{bb}}$
(where $g_{ab}$ is a coupling constant between the components $a$ and $b$). We show
that in the Bose-Einstein-condensation phase ($T\leqslant T_c$) the system may remain in a stable and miscible
phase also for larger values of $g_{ab}$, depending on the gas parameter $\gamma$ and
temperature.

\end{abstract}


\maketitle

\section{Introduction}

The investigation of mixtures of two-component Bose gases has been of interest since the
experimental realization nearly 25 years ago in JILA \cite{myat97,hall98}. Due to the
possibility of tuning the interspecies scattering length ($a_{ab}$) by using Feshbach
resonances, two-component quantum fluids exhibit rich physics that is not accessible in
a single-component fluid. Theoretical \cite{Petrov15, Timmerman} and experimental studies
\cite{papp2008, papp20082, burchanti202, kim20, Lee20, wacher15, wang15, cabrera18, fava18} 
have revealed that the nature of this physics dramatically depends on the sign of the
intercomponent coupling constant of the $s$-wave contact interaction $g_{ab}$: for
$g_{ab}<0$ ($g_{aa}>0$, $g_{bb}>0$), quantum liquid droplets may arise
\cite{Petrov15, cabrera18}, while for $g_{ab}>0$, a phase transition between miscible and
immiscible states can occur. In some sense, the situation is similar to two-body physics:
when the interparticle interaction is negative, one is mainly interested in the properties
of bound states; otherwise one studies scattering angles and cross sections.

In the present work we concentrate only on the case with repulsive interactions;
($g_{aa}>0$, $g_{bb}>0$, $g_{ab}>0$) and theoretically study the properties of a
two-component homogeneous Bose system, such as stability, miscibility, and a possible
phase transition at finite temperature at equilibrium.

Although recent experiments \cite{wang15,Lee20} do not clearly identify evident signatures
of miscible-immiscible transition, the existence of the transition with spatial separation,
including zero temperature, has been theoretically proven \cite{ozaki10, shi2000, schayback13,
roy15, sasaki09, takeuchi13, tickhor13, kim02, hryhorchak, ota20}. Particularly, a long time
ago, Timmermans \ci{Timmerman} proposed to distinguish two types of spatial separation:
(1) potential separation, caused by the external trapping potentials in much the same way
as gravity can separate fluids of different specific weight, and (2) phase separation, which
persists in the absence of external potentials and is similar to separation of immiscible
fluids, such as oil and water. In the present work we discuss the system without a trap and
study only the phase separation which takes  place after crossing the border of instability.
We show that the onset of the instability in the system lowers its free energy by the
segregation of the components into a phase-separated state.

The origin of this instability is the following. In contrast to a single-component Bose
system, a binary mixture of bosons with Bose-Einstein condensation (BEC) has two branches
of collective excitations, $\omega_d$ and $\omega_s$, corresponding to density $c_d$ and
pseudospin sound $c_s$ modes, respectively. The former describes the oscillations of both
components in phase, while the latter is responsible for out-of-phase oscillations of the
components with respect to each other. For some values of physical parameters
($g_{ab}$, $T$) for one of the modes $c_s^2<0$, so that this mode grows initially at
a rate $|\omega_s|$, which is the indicator of instability \cite{Timmerman}. Particularly,
at zero temperature, this may happen when the interspecies coupling constant $g_{ab}$ exceeds
a critical value $g_c=\sqrt{g_{aa}g_{bb}}$, i.e., $g_{ab}>g_c$. Nowadays this criterion
is so widely accepted that some authors consider it even as a definition of miscible
($g_{ab}<\sqrt{g_{aa}g_{bb}}$) or immiscible ($g_{ab}>\sqrt{g_{aa}g_{bb}}$) states
\cite{wacher15,boudjuma} despite the fact that it was obtained in the rather crude
Bogoliubov approximation, which is valid only for very dilute gases, with the gas
parameter $\gamma=\rho a^3\sim 10^{-5}$ \cite{andersen}.

As to the works where some corrections to the Bogoliubov or semiclassical approximations
were considered \cite{ota20,boudjuma,roy15,schayback13,hryhorchak}, they have mainly two
drawbacks: the Hugenholtz-Pines (HP) theorem \cite{pines1959, watabe} for multicomponent BECs
is not satisfied and/or the anomalous densities, especially intercomponent anomalous
pair density, are neglected. As a result, the majority of theoretical approximations lack
self-consistency, being valid only for $\gamma\ll 1$. The principal necessity of taking
into account anomalous averages for the  Bose-condensed phase has been emphasized in Refs.
\cite{Yukalov_PRE,Yukalov_PLA, yukalovannals, ourAniz1, ourannalshJs, mysolo, ourctan2, 
ourAniz2part1,ouraniz2part2, ourpra77}.

Atomic interactions in gases are modeled by contact potentials expressed through
effective scattering lengths, which can be made rather large by means of the Feshbach
resonance technique, so that the gas parameter can become quite large
 \ci{papp2008,navon2011,lopes2017}. The aim of the present
work is to develop a mean-field-based approximation, without any restriction to the value of
$\gamma>0$, by taking into account the HP theorem, derived for multicomponent BECs in Refs.
\cite{watabe, Nepomnyash}, as well as anomalous densities, $\sigma_a$, $\sigma_b$, and
$\sigma_{ab}$. For this purpose we start with the standard Hamiltonian of a binary Bose
mixture with contact interactions. We use the variational method, similar to that
employed in Refs. \cite{ourAniz2part1, pinto, stev81, ouryee04}, which is a variant of
the general approach called optimized perturbation theory \cite{Yukalov_PPN, yukphys21}.
In the present case, this approach is equivalent to the Hartree-Fock-Bogoliubov (HFB)
approximation \cite{yukechaya}.

It is well known that the usage of the HFB approximation has its own problem, which is called
in the literature the Hohenberg-Martin dilemma \cite{HohenbergM}, which is summarized as
follows: In the theory based on the standard grand canonical ensemble with spontaneous
symmetry breaking, depending on the way of calculation, one obtains either a gap in the
spectrum of collective excitations, or local conservation laws together with general
thermodynamic relations becoming invalid. Recall that the excitation spectrum, according
to the HP theorem, must be gapless. A self-consistent way of solving this dilemma has been
advanced in Refs. \cite{Yukalov_PRE,Yukalov_PLA, yukalovannals} by introducing additional 
Lagrange multipliers for each component, namely, $\mu_{0a}$, $\mu_{1a}$, $\mu_{0b}$, and 
$\mu_{1b}$. This choice is directly related to the inclusion of anomalous density. So, when 
one neglects the anomalous density, $\mu_0$ equals  $\mu_1$, while when the anomalous density
is taken into account, $\mu_0$ and $\mu_1$ can be fixed from the conditions of minimization
of the thermodynamic potential with respect to condensed fractions and from the validity
of the generalized HP theorem, respectively \cite{Yukalov_PLA, yukechaya}.

Since there is no restriction to the magnitude of the gas parameter in the present
approach, our criterion for the stability of a binary Bose system will be more general
than the simple inequality $g_{ab}\leqslant \sqrt{g_{aa}g_{bb}}$. Particularly, for a
symmetric system with equal masses ($m_a=m_b$) and coupling constants ($g_{aa}=g_{bb}=g$,
$g_{ab}=\overline{g}_{ab} g$) at zero temperature we have obtained the phase diagram
on the ($\overline{g}_{ab}$, $\gamma$) plane which displays that, for moderate values
of $\gamma$ ($\gamma\approx 0.001$), the system may remain stable even at
$\overline{g}_{ab}=1.1$, in contrast to the predictions of the previous studies.
At finite temperatures this criterion clearly involves at least three parameters
($\overline{g}_{ab}$, $\gamma$, $T$), which gives us the opportunity to derive a
three-dimensional phase diagram for a Bose-condensed two-component homogeneous mixture.

The paper is structured as follows. In Sec. II, we derive general expressions for the
free energy, collective excitation spectrum, and the densities. Then, in Sec. III, we
discuss the BEC system in more detail. The theory is applied to the symmetric Bose mixture
in Sec. IV in order to obtain quantitative results. In the last section, we present
discussions and conclusions.

\section{Thermodynamic potential and main equations}
  The Lagrangian density for two-species complex scalar fields $\psi$ and $\phi$,
with contact self-couplings $g_a$ and $g_b$ and interspecies coupling $g_{ab}$, is
given as
\bea 
\begin{aligned}
&& L=\psi^{\dagger} (i\partial_t+\frac{\vec{\triangledown}^2}{2m_a}+\mu_a)\psi-\frac{g_a}{2}(\psi^{\dagger}\psi)^2+\phi^{\dagger} (i\partial_t \\
&& +\frac{\vec{\triangledown}^2}{2m_b}+\mu_b)\phi-\frac{g_b}{2}(\phi^\dagger\phi)^2-g_{ab}(\psi^\dagger\psi)(\phi^\dagger\phi)
\label{Lagrange}
\end{aligned}
\eea
where the associated chemical potentials are represented by $\mu_{a,b}$ while $m_{a,b}$
represent the masses. In terms of the corresponding $s$-wave scattering lengths $a_s$,
the coupling constants can be written as $g_{a,b}=4\pi a_{a,b}/m_{a,b}$, while the cross
coupling is $g_{ab}=2\pi a_{ab}/m_{ab}$, where $m_{ab}=m_am_b/(m_a+m_b)$ represents reduced
mass. Here and below we set $\hbar=1$, $k_B=1$.

Note that in the present work only repulsive interactions will be considered,
$g_{a,b}\geqslant 0$, $g_{ab}\geqslant 0$. The grand canonical thermodynamic potential
$\Omega$ can be calculated in the path integral formalism as
\be
\Omega=-T\ln Z,
\label{omega}
\ee
\be
Z=\int D\psi^{\dagger}D\psi D\phi^{\dagger}D\phi e^{-S(\psi^{\dagger}, \psi, \phi^{\dagger},
\phi)} \; ,
\label{Z}
\ee
where the equivalent finite-temperature Euclidean $(\tau=it)$ space time action to
(\ref{Lagrange}) is given by
\bea
\begin{aligned}
S=&\int_0^{\beta} d\tau \int d\vec{r} \left\lbrace \psi^{\dagger} \hat{K}_{a}\psi +\phi^{\dagger}\hat{K}_{b}\phi +\frac{g_a}{2}(\psi^{\dagger} \psi)^2\right. \\
&+\left.\frac{g_b}{2}(\phi^{\dagger} \phi)^2+g_{ab}(\psi^{\dagger} \psi)(\phi^{\dagger} \phi)\right\rbrace, \\
 \hat{K}_{a,b}&=\frac{\partial}{\partial\tau}-\hat{O}_{a,b}; \ \ \ \ \ \ \  \ \hat{O}_{a,b}=\frac{\vec{\nabla}^2}{2m_{a,b}}+\mu_{a,b} \; .
\label{action}
\end{aligned}
\eea

In Eq. (\ref{action}) the fields $\psi(\vec{r},\tau)$ and $\phi(\vec{r},\tau)$ are
periodic in $\tau$ with period $\beta=1/T$. Clearly, this path integral cannot be evaluated
exactly due to the terms of fourth order in fields, so  an approximation is needed.
In the present work, we use the approach sometimes called variational perturbation theory
\cite{stevenson, pinto, orkl1, ourkl2}, which is a particular case of optimized perturbation
theory \cite{Yukalov_PPN, yukphys21}. For a two-component system, this method involves
the following steps.

\textit{ Step 1.}	\quad  Introduce fluctuating fields $\tilde{\psi}$ and $\tilde{\phi}$ by the Bogoliubov shift:
\bea
\begin{aligned}
\psi(\vec{r},\tau)=\sqrt{\rho_{0a}}+\tilde{\psi}(\vec{r},\tau) \; , \\
\phi(\vec{r},\tau)=\sqrt{\rho_{0b}}+\Tilde{\phi}(\vec{r},\tau) \; ,
\label{shift}
\end{aligned}
\eea
where the order parameters $\rho_{0a}$ and $\rho_{0b}$ correspond to the condensate
fractions of the components $a$ and $b$, respectively. Note that the Bogoliubov shift is an
exact canonical transformation \cite{Yukalov_LP}, 
and 
not an approximation, as sometimes
it is stated. For a uniform system at equilibrium, $\rho_{0a}$ and $\rho_{0b}$ are real
variational constants, which are fixed by the minimum of the free energy $\Omega$ as
$\partial\Omega/\partial\rho_{0,a,b}=0$,
$\partial^2\Omega/\partial^2\rho_{0,a,b}\geqslant 0$ \cite{ourAniz2part1}.
As to the numbers of uncondensed particles $N_{1a}$ and $N_{1b}$, they are related to
the fields $\tilde{\psi}$ and $\tilde{\phi}$:
\bea
&N_{1a}=V\rho_{1a}=\int d\vec{r}\langle\tilde{\psi}^\dagger(r)\tilde{\psi}(r)\rangle, \\
&N_{1b}=V\rho_{1b}=\int d\vec{r}\langle\tilde{\phi}^\dagger(r)\tilde{\phi}(r)\rangle \; ,
\label{n1a1b}
\eea
so that
\bea
&&N_a=\int d\vec{r}\langle \psi^\dagger(r)\psi(r) \rangle, \\
&&N_b=\int d\vec{r}\langle \phi^\dagger(r)\phi(r)
\label{n1a1b1}
\rangle
\eea
with the normalization conditions $N=N_a+N_b$, $N_a=V\rho_a=V(\rho_{0a}+\rho_{1a})$,
and $N_b=V\rho_b=V(\rho_{0b}+\rho_{1b})$, where $N_{a(b)}$ is the number of particles in the
component $a$, ($b$) and $N$ is the particle number in the whole two-component system 
and $V$ is the total volume of the system.
Since we are considering a homogeneous system, the densities $\rho_a$ and $\rho_b$ are
uniform.

\textit{Step 2.} \quad Make the following replacement in the action:
$g_a \rightarrow {\tilde\delta} g_a$, $g_b \rightarrow {\tilde\delta} g_b$,
$g_{ab} \rightarrow {\tilde\delta} g_{ab}$.

\textit{Step 3.} \quad Add to the action the term:
\bea
\begin{aligned}
S_\Sigma=&(1-{\tilde\delta})\int d\tau d\vec{r}\Big\{ \Sigma_{n}^{(a)}(\tilde{\psi}^\dagger \tilde{\psi})+ \Sigma_{n}^{(b)}(\tilde{\phi}^\dagger \tilde{\phi}) \\
&+\frac{1}{2}\Sigma_{an}^{(a)}(\tilde{\psi}\tilde{\psi}+\tilde{\psi}^\dagger\tilde{\psi}^\dagger)+ \frac{1}{2}\Sigma_{an}^{(b)}(\tilde{\phi}\tilde{\phi}+\tilde{\phi}^\dagger\tilde{\phi}^\dagger) \\
&+\Sigma_{n}^{(ab)}(\tilde{\psi}^\dagger\tilde{\phi}+\tilde{\phi}^\dagger\tilde{\psi})+\Sigma_{an}^{(ab)}(\tilde{\psi}\tilde{\phi}+\tilde{\phi}^\dagger\tilde{\psi}^\dagger)\Big\} \; ,
\end{aligned}
\eea
where the variational parameters $\Sigma_n$ and $\Sigma_{an}$ can be naturally interpreted
as normal and anomalous self-energies, respectively.

\textit{Step 4.} \quad Now in the Cartesian representation
\bea
\begin{aligned}
\tilde{\psi}=\frac{1}{\sqrt{2}}(\psi_1+i\psi_2), \  \ \ \ \ \ \tilde{\phi}=\frac{1}{\sqrt{2}}(\psi_3+i\psi_4) \; ,
\end{aligned}
\lab{dekart}
\eea
such that
\bea
\int D\tilde{\psi}^\dagger D\tilde{\psi}D\tilde{\phi}^\dagger D\tilde{\phi} \rightarrow \int \prod_{i=1}^4 D\psi_i \; ,
\label{psi12}
\eea
the action (\ref{action}) can be written as
\bea
\label{3.1}
S=&&S_0+S_{free}+{\tilde \delta}S_{int}, \\
S_{int}&&=S^{(1)}_{int}+S^{(2)}_{int}+S^{(3)}_{int}+S^{(4)}_{int}, \\
S_0=&&\int_0^\beta d\tau\int d\vec{r}\left\{ -\mu_{0a}\rho_{0a}-\mu_{0b}\rho_{0b}+\frac{g_a\rho^2_{0a}}{2}\right. \nonumber\\
&&\left.+\frac{g_b\rho^2_{0b}}{2}+g_{ab}\rho_{0a}\rho_{0b}\right\}, \\
\label{3.2}
S_{free}&&=\frac{1}{2}\int_0^\beta d\tau\int d\vec{r}\lbrace \psi_1[X_1+\hat{K}_a]\psi_1+ \psi_2[X_2+\hat{K}_a]\psi_2 \nonumber\\
&&+\psi_3[X_3+\hat{K}_b]\psi_3+ \psi_4[X_4+\hat{K}_b]\psi_4+X_5[\psi_1\psi_3 \nonumber\\
&&+\psi_3\psi_1]+X_6 [ \psi_2\psi_4+\psi_4\psi_2]\rbrace, 
\eea
\bea
\label{3.3}
S_{int}^{(2)}=&&\frac{1}{2}\int d\tau d\vec{r}\left\lbrace \sum_{i=1}^4\Lambda_i\psi^2_i+\Lambda_{5}(\psi_1\psi_3+\psi_3\psi_1)\right. +\Lambda_{6}(\psi_2\psi_4+\psi_4\psi_2) \Bigg\}, \\
S_{int}^{(4)}=&&\frac{1}{8}\int d\tau d\vec{r}\Big\{ g_a(\psi_1^2+\psi_2^2)^2+g_b(\psi_3^2+\psi_4^2)^2+2g_{ab}(\psi_1^2+\psi_2^2)(\psi_3^2+\psi_4^2) \Big\},
\eea
where we have introduced the following notations:
\bea
&&X_{1}=\Sigma_n^{(a)}+\Sigma_{an}^{(a)}-\mu_{1a},   X_{3}=\Sigma_n^{(b)}+\Sigma_{an}^{(b)}-\mu_{1b},
\label{4.1}\ \  \\
&&X_{2}=\Sigma_n^{(a)}-\Sigma_{an}^{(a)}-\mu_{1a},  X_{4}=\Sigma_n^{(b)}-\Sigma_{an}^{(b)}-\mu_{1b},
\label{4.2} \ \ \\
&&X_{5}=\Sigma_n^{(ab)}+\Sigma_{an}^{(ab)}, \qquad X_{6}=\Sigma_n^{(ab)}-\Sigma_{an}^{(ab)},
\label{4.3}\\
&&\Lambda_1=-\mu_{1a}-X_1+3g_a\rho_{0a}+g_{ab}\rho_{0b},
\label{4.4}\\
&&\Lambda_2=-\mu_{1a}-X_2+g_a\rho_{0a}+g_{ab}\rho_{0b},
\label{4.5}\\
&&\Lambda_3=-\mu_{1b}-X_3+3g_b\rho_{0b}+g_{ab}\rho_{0a},
\label{4.6}\\
&&\Lambda_4=-\mu_{1b}-X_4+g_b\rho_{0b}+g_{ab}\rho_{0a},
\label{4.7}\\
&&\Lambda_{5}=-X_{5}+2g_{ab}\sqrt{\rho_{0a}\rho_{0b}},
\label{4.8}\\
&&\Lambda_{6}=-X_{6},
\label{4.9}
\eea
Equations (\ref{3.1}) -- (\ref{4.9}) need some comments:
\begin{enumerate}
\item[a)] We omit the explicit expressions for $S^{(1)}$ and $S^{(3)}$ since the path
integrals including odd powers of fields are zero, e.g.,
$\int [ \prod_{i=1}^{4}D\psi_i ]\psi_1\psi_2^2 e^{-S}=0$.
\item[b)] Two kinds of chemical potentials, $\mu_0$ and $\mu_1$, are introduced instead
of a unique chemical potential $\mu$, such that $\mu_{0a}N_{0a}+\mu_{1a}N_{1a}=\mu_{a}N_{a}$.
The reason is the following. Actually, the mean-field-based theories of BEC have a
long-standing puzzle referred to as the Hohenberg-Martin dilemma \cite{HohenbergM}, which
can be explained for a one-component system rather simply. The chemical potential should
satisfy the Goldstone theorem and correspond to the minimum of the thermodynamic potential.
It has been shown that, when anomalous density, 
$\sigma\sim\langle\tilde{\psi}\tilde{\psi}\rangle$\footnote{
See Eq. \re{SigaSigb} below.
}
is accurately taken into account, these
two conditions cannot be satisfied simultaneously \cite{Yukalov_PLA, yukalovannals, ourAniz1}.
The solution to this problem has been advanced in Refs. \cite{Yukalov_PRE, Yukalov_PLA, yukalovannals}.
It was shown that, in a system with spontaneous gauge symmetry breaking, the introduction
of two chemical potentials makes the theory self-consistent. Naturally, in the normal phase,
when $\rho_0=0$, $\sigma=0$, both chemical potentials coincide: $\mu=\mu_0=\mu_1$.
\item[c)] The six variational parameters $X_1,\ldots,X_6$ should be fixed by the minimization
of $\Omega$, e.g., $\partial\Omega/(\partial X_{i})=0$, ($i=1--6$)
\end{enumerate}
\textit{Step 5.} \quad Now, passing to the momentum space,
\be
{\psi}_i(\vec{r},\tau)=\frac{1}{\sqrt{V\beta}}\sum_{n=-\infty}^\infty \sum_k {\psi}_i(\omega_n,\vec{k})e^{i\omega_n\tau+i\vec{k}\vec{r}},
\ee
where $\sum_{k}=V\int d\vec{k}/(2\pi)^3$ and $\omega_n=2\pi n T$ is a Matsubara frequency,
one may present Eq. (\ref{3.2}) as
\bea
S_{free}=&&\frac{(2\pi)^4}{2V\beta}\sum_{\vec{k},\vec{p},m,n}\sum_{i,j=1}^4\psi_i(\omega_n,\vec{k})G_{ij}^{-1}(\omega_n,\vec{k};\omega_m,\vec{p}) \nonumber \\
&&\times\psi_j(\omega_m,\vec{p})\delta(\vec{k}+\vec{p})\delta(\omega_n+\omega_m)
\eea
with the inverse propagator
\be
G^{-1}(\omega_n,\vec{k})=
\begin{pmatrix}
\veps_a(k)+X_1 & \omega_n & X_5 & 0 \\
-\omega_n & \veps_a(k)+X_2 & 0 & X_6 \\
X_5 & 0 & \veps_b(k)+X_3 & \omega_n \\
0 & X_6 & -\omega_n & \veps_b(k)+X_4
\end{pmatrix},
\label{greeninv}
\ee
where $\veps_{a,b}(k)=\vec{k}^2/{2m_{a,b}}$ \; .

Evaluating the determinant of this matrix, we obtain two branches of dispersion:
\bea
\omega_{1,2}(k)&&=\sqrt{\frac{E_a^2+E_b^2}{2}+X_5 X_6\pm \frac{\sqrt{D}}{2}}
\label{dispX5X6}, \\
D=&&(E_a^2-E_b^2)^2+4E_{13}^2X_6^2+4E_{24}^2X_5^2\nonumber \\
&&+4X_5X_6(E_a^2+E_b^2)\; ,
\eea
where
\bea
&E_a^2=(\veps_a(k)+X_1)(\veps_a(k)+X_2), \\ &E_b^2=(\veps_b(k)+X_3)(\veps_b(k)+X_4),\\
&E_{13}^2=(\veps_a(k)+X_1)(\veps_b(k)+X_3), \\ &E_{24}^2=(\veps_a(k)+X_2)(\veps_b(k)+X_4) \; .
\lab{Eab}
\eea

\textit{Step 6.} \quad The perturbation scheme is considered as an expansion in powers of ${\tilde\delta}$
by using the propagators
\be
G_{ij}(r,\tau;r^\prime, \tau^\prime)=\frac{1}{V\beta}\sum_{n,\vec{k}}e^{i\vec{k}(r-r^\prime)}e^{i\omega_n(\tau-\tau^\prime)}G_{ij}(\omega_n,\vec{k}) \; ,
\ee
which are presented explicitly in the Appendix. The expansion parameter ${\tilde\delta}$ will
be set to ${\tilde\delta}=1$ at the end of the calculations.

\textit{Step 7.} \quad The detailed calculation of the generating functional and hence $\Omega$ in the first
order of $\tilde\delta$ can be performed in the similar way as it has been done in
Ref. \cite{ourAniz2part1} for the one-component model. Therefore, using the following
formulas, where $(x=(\tau,\vec{r}))$,
\bea
&&\langle \psi_i(x)\psi_j(x)\rangle=G_{ij}(x,x)=\frac{1}{V\beta}\sum_{\omega_n,\vec{k}}G_{ij}(\omega_n,\vec{k}), \\
&&\langle \psi_i^2(x)\psi_j^2(x)\rangle=G_{ii}(x,x)G_{jj}(x,x)+2G_{ij}^2(x,x),  \ \ \\
&&\langle \psi_i^4(x)\rangle=3G_{ii}^2(x,x), \ \ \ \ \ \ G_{ij}(x,x)=G_{ji}(x,x), \ \ \\
&&G_{12}(x,x)=G_{14}(x,x)=0, \\
&& G_{23}(x,x)=G_{34}(x,x)=0 ,
\eea
one obtains
\bea
\begin{aligned}
 \Omega=&\Omega_0+\Omega_{ln}+\Omega_2+\Omega_4 \\
 \Omega_0&=V\left\lbrace -\mu_{0a}\rho_{0a}-\mu_{0b}\rho_{0b}+\frac{g_a\rho_{0a}^2}{2}+\frac{g_b\rho_{0b}^2}{2}+g_{ab}\rho_{0a}\rho_{0b} \right\rbrace \\
 \Omega_{ln}&=\frac{T}{2}\sum_{k,\omega_n} \ln [(\omega_n^2+\omega_1^2)(\omega_n^2+\omega_2^2)]=\frac{1}{2}\sum_k(\omega_1(k)+\omega_2(k))+\\
&+T\sum_k \ln(1-e^{-\beta\omega_1(k)})+T\sum_k \ln(1-e^{-\beta\omega_2(k)}), \\
 \Omega_2&=\frac{1}{2}\sum_{i=1}^{6} A_i \Lambda_i, \\
 \Omega_4&=\frac{1}{8V}\Bigg\{ g_a[3A_1^2+3A_2^2+2A_1A_2]+g_b[3A_3^2+3A_4^2+2A_3A_4] \\
& +2g_{ab}\left[(A_1+A_2)(A_3+A_4)+\frac{A_5^2+A_6^2}{2}\right]\Bigg\},
\label{omegbig}
\end{aligned}
\eea
where $A_i=VG_{ii}(x,x) \ (i=1 - 4)$, $A_5=2VG_{13}(x,x)$, $A_6=2VG_{24}(x,x)$,
$\Lambda_i$ are given by Eqs. (\ref{4.4})--(\ref{4.9}), and $G_{ij}(x,x)$ are presented
in the Appendix. 
Feynman diagrams contributing to $\Omega$ in the present optimized perturbation theory 
are illustrated in Refs. \ci{andersen,stev42}.\footnote{
The next order corrections to the present approximation can be found in the similar
way as has been developed by Stancu and Stevenson \ci{stev42} for the simple $\lambda\phi^4$ 
theory.
}

The variational parameters are determined  by the minimization of the thermodynamic
potential $\Omega(X_1,\ldots,X_6)$ as $\partial\Omega(X_1,\ldots,X_6)/\partial X_i=0 \ \ (i=1-6)$.
These equations can be rewritten in the following compact form:
\bea
\begin{aligned}
&X_1=g_a[3\rho_{0a}+2\rho_{1a}+\sigma_a]+g_{ab}\rho_b-\mu_{1a}, \\
&X_2=g_a[\rho_{0a}+2\rho_{1a}-\sigma_a]+g_{ab}\rho_b-\mu_{1a}, \\
&X_3=g_b[3\rho_{0b}+2\rho_{1b}+\sigma_b]+g_{ab}\rho_a-\mu_{1b}, \\
&X_4=g_b[\rho_{0b}+2\rho_{1b}-\sigma_b]+g_{ab}\rho_a-\mu_{1b}, \\
&X_5=2g_{ab}\sqrt{\rho_{0a}\rho_{0b}}+g_{ab}\frac{\rho_{ab}+\sigma_{ab}}{2}, \\
&X_6=\frac{g_{ab}}{2}(\rho_{ab}-\sigma_{ab}) \; ,
\label{EqsX16}
\end{aligned}
\eea
where the densities $\rho_1$ and $\sigma$ are given in the next section. Note that,
in the derivation of Eqs. (\ref{EqsX16}), we used the relation
$\partial\Omega_{ln} / \partial X_i=A_i/2$,  $i=1- 6$, which can be checked by using
\textit{Mathematica} or MAPLE. In general, the system of Eqs. \re{EqsX16} with the given set of 
input parameters, such as coupling parameters, and the total densities of atoms is the system  
of nonlinear algebraic equations with respect to unknown variational parameters  
$(X_1,\ldots,X_6)$. As it is seen from their definition in Eqs. \re{4.1}--\re{4.9}  the latter can 
be clearly considered  self-energies in the Cartesian representation \re{dekart}.

\subsection{Normal and anomalous densities}

Fluctuating fields $\Tilde{\psi}(r)$ and $\Tilde{\phi}(r)$ define the density of
uncondensed particles in accordance with Eqs. (\ref{n1a1b}). When the Green's functions
are known, these densities may be calculated as
\bea
\begin{aligned}
\rho_{1a}=&\frac{1}{V}\int d\vec{r}\langle \Tilde{\psi}^\dagger(\vec{r})\Tilde{\psi}(\vec{r})\rangle=\frac{1}{2V}\int d\vec{r}[G_{11}(\vec{r},\vec{r})\\
&+G_{22}(\vec{r},\vec{r})]=\frac{1}{2V}\left(A_1+A_2\right),\\
\rho_{1b}=&\frac{1}{V}\int d\vec{r}\langle \Tilde{\phi}^\dagger(\vec{r})\Tilde{\phi}(\vec{r})\rangle=\frac{1}{2V}\int d\vec{r}[G_{33}(\vec{r},\vec{r})\\
&+G_{44}(\vec{r},\vec{r})]=\frac{1}{2V}(A_3+A_4).
\label{rho1a.rho1b}
\end{aligned}
\eea
In general, one may introduce the anomalous
\bea
\begin{aligned}
\sigma_a & =\frac{1}{2V} \int d\vec{r}[\langle \Tilde{\psi}^\dagger(\vec{r})\Tilde{\psi}^\dagger(\vec{r})\rangle + \langle \Tilde{\psi}(\vec{r})\Tilde{\psi}(\vec{r})\rangle]\\
& =\frac{1}{2V} (A_1-A_2),\\
\sigma_b & =\frac{1}{2V} \int d\vec{r}[\langle \Tilde{\phi}^\dagger(\vec{r})\Tilde{\phi}^\dagger(\vec{r})\rangle + \langle \Tilde{\phi}(\vec{r})\Tilde{\phi}(\vec{r})\rangle] \\
& =\frac{1}{2V} (A_3-A_4)
\label{SigaSigb}
\end{aligned}
\eea
and ``mixed" densities:
\bea
\begin{aligned}
\rho_{ab} & =\frac{1}{V} \int d\vec{r}[\langle \Tilde{\psi}^\dagger(\vec{r})\Tilde{\phi}(\vec{r})\rangle + \langle \Tilde{\phi}^\dagger(\vec{r})\Tilde{\psi}(\vec{r})\rangle] \\
 &=\frac{1}{V}\int d\vec{r}[G_{13}(\vec{r},\vec{r})+G_{24}(\vec{r},\vec{r})]=\frac{1}{2V} (A_5+A_6),  \\
 \sigma_{ab}&=\frac{1}{V} \int d\vec{r}[\langle \Tilde{\psi}(\vec{r})\Tilde{\phi}(\vec{r})\rangle + \langle \Tilde{\phi}^\dagger(\vec{r})\Tilde{\psi}^\dagger(\vec{r})\rangle]\\
 &= \frac{1}{2V} (A_5-A_6).
\label{RhoSigab}
\end{aligned}
\eea
Clearly, these densities, which are explicitly given in the Appendix, do not depend on the
coordinate variables; i.e., they are constants for a uniform system. Physically, the pair
densities $\rho_{ab}$ and $\sigma_{ab}$ describe the processes where, due to the presence
of the reservoir, particles are exchanged or pairing correlations emerge between the two
components.

From their definition, it is clear that the mixed densities characterize the correlations
between the components of a two-component system. To quantify these correlations, one may
introduce the overlap parameter $\eta$,
\be
\eta=\frac{1}{2\sqrt{N_aN_b}}\int d\vec{r}\lbrace \langle \psi^\dagger(\vec{r})\phi(\vec{r})\rangle+\langle \phi^\dagger(\vec{r})\psi(\vec{r})\rangle \rbrace
\label{etadef}
\ee
where $\psi(\vec{r})$ and $\phi(\vec{r})$ are the field operators of the components $a$
and $b$, respectively.

Using Eqs. (\ref{shift}), (\ref{dekart}), and (\ref{EqsX16}) -- \re{etadef}, one may present
$\eta$ as follows:
\bea
\eta=&&\frac{1}{\sqrt{N_a N_b}}\int d\vec{r}\sqrt{\rho_{0a}\rho_{0b}}+\frac{1}{2\sqrt{N_a N_b}}\int d\vec{r} \lbrace\langle\Tilde{\psi}^{\dagger}(\vec{r})\Tilde{\phi}(\vec{r})\rangle \nonumber\\
&&+\langle\Tilde{\phi}^{\dagger}(\vec{r})\Tilde{\psi}(\vec{r})\rangle\rbrace=\sqrt{n_{0a} n_{0b}}+\frac{\rho_{ab}}{2\sqrt{\rho_a\rho_b}}=\frac{X_5+X_6}{2g_{ab} \sqrt{\rho_a\rho_b} } \; ,\nonumber\\
\label{etaRhoab}
\eea
where $n_{0a}$ and $n_{0b}$ are the normalized condensed fractions, $n_{0a}=\rho_{0a}/\rho_{a}$,
$n_{0b}=\rho_{0b}/\rho_{b}$. Note that, when the fluctuations are neglected, i.e.,
$\Tilde{\psi}=\Tilde{\phi}=0$, $\eta$ in Eq. (\ref{etaRhoab}) coincides with the miscibility
parameter of Refs. \cite{etawenfa,dipolarbrazileta}, introduced for nonuniform coupled
systems. Particularly, when at least one of the components is in the normal phase, the parameter $\eta$
is completely defined by the normal pair density,
$\eta (T>T_c)=\rho_{ab}/2\sqrt{\rho_a\rho_b}=X_5/g_{ab} \sqrt{\rho_a\rho_b}$.

\subsection{Particular cases of HFB approximation}

The presented HFB-type theory is general, so that some well-known approximations to this 
general theory can be easily derived as particular cases.

i) Sometimes, one uses the trick (first suggested by Shohno \cite{Shohno}) of omitting 
anomalous averages, which corresponds to the case when in 
Eqs. \re{omegbig} and  \re{EqsX16} $\sigma_a$, $\sigma_b$ , and  $\sigma_{ab}$ are omitted by 
setting $\mu_{0a,b}=\mu_{1a,b}=\mu_{a,b}$. However, as has been shown in a number of 
publications \cite{Yukalov_PRE,Yukalov_PLA, yukalovannals, ourAniz1, ourannalshJs, mysolo, 
ourctan2, ourAniz2part1,ouraniz2part2, ourpra77}, this trick results in a not-self-consistent
approach containing paradoxes. 

ii) Bogoliubov and quadratic approximations correspond to the case when, after the shift 
\re{shift} only quadratic terms of fluctuating fields are kept in the action
\re{action} : $S\approx S_0+S_{free}+S_{int}^{(2)}$, with $\tilde\delta=1$. The formal 
difference is that in the quadratic approximation one has
\be
\Omega^{Bil}=\Omega_0+\Omega_{ln},
\lab{ombil}
\ee
where $\Omega_0$ and $\Omega_{ln}$ have the same expressions as in Eqs. \re{omegbig} with the 
self-energies given by
\bea
\begin{aligned}
&X_1\approx X_1^{Bil}=3g_a\rho_{0a}+g_{ab}\rho_{0b}-\mu_{a}, \\
&X_2\approx X_2^{Bil}=g_a\rho_{0a}+g_{ab}\rho_{0b}-\mu_{a}, \\
&X_3\approx X_3^{Bil}= 3g_b\rho_{0b}+g_{ab}\rho_{0a}-\mu_{b},\\
&X_4\approx X_4^{Bil}= g_b\rho_{0b}+g_{ab}\rho_{0a}-\mu_{b},\\
&X_5\approx X_5^{Bil}=2g_{ab}\sqrt{\rho_{0a}\rho_{0b}}, \\
&X_6\approx X_6^{Bil}=0. 
\label{EqsX16_BIL}
\end{aligned}
\eea 

In the Bogoliubov approximation, the thermodynamic potential is formally given by 
Eq.\re{ombil}, and  the self-energies by Eqs. \re{EqsX16_BIL}, with setting there 
$\rho_{0a,b}\approx \rho_{a,b}$ , i.e., 
$X_{i}^{Bog}=X_{i}^{Bil}(\rho_{0a}=\rho_{a},\rho_{0b}=\rho_{b})$. In this case, 
the equations are uncoupled and the solutions are simple. Actually, both these variants
enjoy the same level of accuracy and are valid only for small gas parameters 
$\gamma \le 10^{-5}$. 

Note that, for all above cases the expressions for the energy dispersions as well as
for the densities are formally the same as given by Eqs. \re{dispX5X6} and 
\re{App_rho} -- \re{App_sig}, respectively.

\subsection{Intermediate summary}
Now we are in a position of summarizing the present section. In practical calculations, in the framework of our
self-consistent theory,
one has to solve the system of, in general,  six nonlinear algebraic equations (\ref{EqsX16}) with
respect to the variational parameters $[X_1,\ldots,X_6]$ and then evaluate all thermodynamic
equilibrium characteristics of the uniform two - component Bose system from
$\Omega(X_1,\ldots,X_6)$ given in (\ref{omegbig}). Stability and miscibility properties can
be studied by analyzing the spectrum of collective excitations (\ref{dispX5X6}).
At a first glance, this procedure, especially, solving the system of six nonlinear algebraic
equations, seems rather cumbersome. However, in reality the number of the unknown variational
parameters $[X_1,\cdots,X_6]$ may be reduced depending on the considered state (BEC or normal
phase) and on the existing symmetries in the system. In the next sections we discuss these
cases in detail.

\section{Condensed and normal phases}

\subsection{Condensed phase}

In this phase, the number of variational parameters is reduced due to the Hugenholtz-Pines
theorem \ci{pines1959}, which has been extended for multicomponent Bose-Einstein
condensates in Refs. \ci{Nepomnyash, watabe}. For a two-component Bose system, in our notation,
it reads
\bea
\begin{aligned}
&\Sigma_n^{(a)}-\Sigma_{an}^{(a)}=\mu_{1a}, \ \ \ \ \ \ \Sigma_n^{(b)}-\Sigma_{an}^{(b)}=\mu_{1b}, \\
&\Sigma_n^{(ab)}=\Sigma_{an}^{(ab)},
\label{pines}
\end{aligned}
\eea
and hence
\bea
\begin{aligned}
X_2=0, \ \ \ \ \ X_4=0, \ \ \ X_6=0, \ \ \ \sigma_{ab}=\rho_{ab} \; .
\label{pinesx}
\end{aligned}
\eea
Therefore, in the BEC phase, instead of six equations, we are left with a system of three
equations:
\be
\left\lbrace\begin{split}
& \Delta_a\equiv X_1/2=g_a(\rho_{0a}+\sigma_{a})=g_a[\rho_a-\rho_{1a}+\sigma_a] \\
& \Delta_b\equiv X_3/2=g_b(\rho_{0b}+\sigma_{b})=g_b[\rho_b-\rho_{1b}+\sigma_b] \\
& \Delta_{ab}\equiv X_5/2=g_{ab}\sqrt{\rho_{0a}\rho_{0b}}+\frac{g_{ab}}{2}\rho_{ab}  \; .
\end{split}\right.
\label{Eqdelta}
\ee
Note that, for the Bose systems with fixed chemical potentials \cite{ourctan2}, these
equations may be rewritten as
\be
\left\lbrace\begin{split}
& \Delta_a=\mu_{1a}+2g_a(\sigma_a-\rho_{1a})-g_{ab}\rho_b \\
& \Delta_b=\mu_{1b}+2g_b(\sigma_b-\rho_{1b})-g_{ab}\rho_a \\
& \Delta_{ab}=g_{ab}(\sqrt{\rho_{0a}\rho_{0b}}+\rho_{ab}/2)  \; .
\end{split}\right. \label{DeltaSimp}
\ee
On the other hand, if the densities are fixed, as in atomic gases, one may determine
the chemical potentials from Eqs. (\ref{EqsX16}) and (\ref{pines}) as
\bea
\begin{aligned}
&\mu_{1a}=g_a[\rho_a+\rho_{1a}-\sigma_a]+g_{ab}\rho_b,  \\
&\mu_{1b}=g_b[\rho_b+\rho_{1b}-\sigma_b]+g_{ab}\rho_a,  \; .
\label{mu1BEC}
\end{aligned}
\eea

The total chemical potentials defined as $\mu_a=(\partial F/\partial N_a)$ and
$\mu_b=(\partial F/\partial N_b)$ (where $F$ is the total free energy of the system) can
be calculated as
\bea
\begin{aligned}
\mu_a\rho_a=\mu_{1a}\rho_{1a}+\mu_{0a}\rho_{0a},
\mu_b\rho_b=\mu_{1b}\rho_{1b}+\mu_{0b}\rho_{0b} \; ,
\label{muabtotalBEC}
\end{aligned}
\eea
where
\bea
\begin{aligned}
& \mu_{0a}=g_a[\rho_a+\rho_{1a}+\sigma_a]+g_{ab}\left[\rho_b+\frac{\rho_{0b}\sigma_{ab}}{\sqrt{\rho_{0a}\rho_{0b}}}\right], \\
& \mu_{0b}=g_b[\rho_b+\rho_{1b}+\sigma_b]+g_{ab}\left[\rho_a+\frac{\rho_{0a}\sigma_{ab}}{\sqrt{\rho_{0a}\rho_{0b}}}\right] \; .
\label{mu0BEC}
\end{aligned}
\eea
The last two equations are derived from $\partial\Omega/\partial\rho_{0a}=0$ and
$\partial\Omega/\partial\rho_{0b}=0$, where $\Omega$ is given by Eq. (\ref{omegbig}). As is
expected, when one neglects anomalous densities by setting $\sigma_a=\sigma_b=\sigma_{ab}=0$,
then $\mu_{0a}=\mu_{1a}=\mu_a$, $\mu_{0b}=\mu_{1b}=\mu_b$.

With the constraints (\ref{pinesx}), the dispersions in Eqs. (\ref{dispX5X6}) can be rewritten
as
\bea
\begin{aligned}
& \omega_{1,2}=\sqrt{\frac{\veps(k)^2}{2}(\nu_1^2+\nu_2^2)+2\veps(k)\Lambda_{1,2}}, \\
& \Lambda_{1,2}=\frac{1}{2}(\Delta_a\nu_1+\Delta_b\nu_2)\pm\frac{\sqrt{D_s}}{4}, \\
& D_s=16\nu_1\nu_2\Delta_{ab}^2+(\nu_1^2\veps(k)-\nu_2^2\veps(k)+2\Delta_a\nu_1-2\Delta_b\nu_2)^2,\\
& \omega_1^2-\omega_2^2=\veps(k)\sqrt{D_s} \; ,
\label{dispBEC}
\end{aligned}
\eea
where we introduce the reduced mass $m_R=m_{ab}=m_am_b/(m_a+m_b)$, and $\nu_1=m_b/(m_a+m_b)$,
$\nu_2=m_a/(m_a+m_b)$, and $\veps(k)=\vec{k}^2/2m_R$. Decomposing $\omega_i$ in powers of
momenta gives the sound velocities through the equations
\bea
\begin{aligned}
\omega_1=c_1|\vec{k}|+O(k^3) \quad,  \quad \omega_2=c_2|\vec{k}|+O(k^3) \;,
\label{omega12}
\end{aligned}
\eea
\bea
\begin{aligned}
&c_1^2=\frac{\Delta_am_b+\Delta_bm_a+\sqrt{4m_am_b\Delta_{ab}^2+(\Delta_b m_a-\Delta_a m_b)^2}}{2m_am_b} \; , \\
\\
&c_2^2=\frac{\Delta_am_b+\Delta_bm_a-\sqrt{4m_am_b\Delta_{ab}^2+(\Delta_b m_a-\Delta_a m_b)^2}}{2m_am_b} \; .
\label{soundBEC}
\end{aligned}
\eea
The velocities $c_1$ ($c_2$) are referred in the literature as density (pseudospin) sound
velocities. In a binary superfluid, the density sound corresponds to oscillation of
two superfluid components in phase, while the pseudospin sound corresponds to the out-of-phase
oscillations \cite{kim20}.

From the last equation it is seen that, when $\Delta_a\Delta_b<\Delta^2_{ab}$, $c_2^2$
becomes negative, signaling the instability of the system. In particular, applying the
Bogoliubov approximation, i.e., setting $\rho_{0a}\approx\rho_a$, $\rho_{0b}\approx\rho_b$, and
$\sigma_a=\sigma_b=\rho_{ab}\approx 0$, then from Eqs. (\ref{Eqdelta}) one obtains
\bea
\begin{aligned}
\Delta_a\approx g_a\rho_a \; , \ \ \ \
\Delta_b\approx g_b\rho_b \; , \ \ \
\Delta_{ab}\approx g_{ab}\sqrt{\rho_a \rho_b} \; ,
\label{DeltaBog}
\end{aligned}
\eea
thus arriving at the well-known stability condition $g_ag_b/g_{ab}^2\geqslant 1$.

Explicit expressions for the densities may be obtained from Eqs.
(\ref{rho1a.rho1b})--(\ref{RhoSigab}) by setting there $X_2=X_4=X_6=0$. As is expected, when
the intercoupling constant goes to zero, we arrive at the well-known formulas of the
single-component case:
\bea
\begin{aligned}
& \rho_{1a}(g_{ab}\rightarrow 0)=\frac{1}{V}\sum_k\left[\frac{\Delta_a+\veps_{a} (\vec k) }{\omega_a(k)}W_1(k)-\frac{1}{2} \right] \; , \\
& \rho_{1b}(g_{ab}\rightarrow 0)=\frac{1}{V}\sum_k\left[\frac{\Delta_b+\veps_{b} (\vec k)}{\omega_b(k)}W_2(k)-\frac{1}{2} \right] \; , \\
& \sigma_a(g_{ab}\rightarrow 0)=-\frac{\Delta_a}{V}\sum_k\frac{W_1(k)}{\omega_a(k)} \; , \\
& \sigma_b(g_{ab}\rightarrow 0)=-\frac{\Delta_b}{V}\sum_k\frac{W_2(k)}{\omega_b(k)} \; , \\
& \rho_{ab}(g_{ab}\rightarrow 0)=\sigma_{ab}(g_{ab}\rightarrow 0)=0 \; , \\
& c_1^2=\Delta_b/m_b \; , \ \ \ \ \ \ \ c_2^2=\Delta_a/m_a \; ,
\label{denseBECg120}
\end{aligned}
\eea
where $W_{a,b}(k)=1/2+1/(e^{\omega_{a,b}(k)\beta}-1)$,
$\omega_{a,b}=\sqrt{\veps_{a,b}(k)(\veps_{a,b}(k)+2\Delta_{a,b})}$ \; .

In the above discussion, we have assumed that both components are in the BEC state. In the
next section, we consider the case where the whole system is in the normal phase; hence
there is no the HP theorem.

\subsection{Normal phase}

The general criterion of miscibility or immiscibility is prescribed by the behavior of the
spectrum of collective excitations. To be miscible, a binary Bose mixture has to possess
all real branches of the collective spectrum positive (non-negative). In the case of a 
Bose-condensed system, the global gauge symmetry is broken and the spectra of single-particle 
and collective excitations coincide \ci{Gavoret}. However, for a normal (uncondensed) system, 
these spectra are different. The positiveness of the single-particle spectrum, defined by 
the poles of the single-particle Green's function, tells us that, on the level of single-particle 
properties, the system is stable, but this tells us nothing about whether it is mixed or 
separated. The spectrum of collective excitations of a normal system is defined by the poles 
of the second-order Green's function or the poles of dynamic susceptibility (response function). 
The mixture separates when the lowest branch of the collective spectrum crosses zero.

First, let us prove that the binary normal Bose mixture at $T > T_c$ enjoys a stable
single-particle spectrum that is positive at any temperature and the values of local
interaction parameters, independently of whether the system is mixed or separated.

By definition, in the normal phase $\rho_{0a}=\rho_{0b}=\sigma_a=\sigma_b=\sigma_{ab}=0$,
and hence $\rho_{1a}=\rho_a$,  $\rho_{1b}=\rho_b$, $\mu_{0a}=\mu_{1a}=\mu_a$,
$\mu_{0b}=\mu_{1b}=\mu_b$, $X_6=X_5$, $X_2=X_1$, and $X_4=X_3$. Thus, the main Eqs. (\ref{EqsX16})
are simplified as
\bea
\begin{aligned}
& X_1=2\rho_a g_a+\rho_b g_{ab}-\mu_a\equiv -\mu_{\rm eff}^{(a)} \; , \\
& X_3=2\rho_b g_b+\rho_a g_{ab}-\mu_b\equiv -\mu_{\rm eff}^{(b)} \; , \\
& X_5=\frac{1}{2}\rho_{ab}g_{ab} \; ,
\label{EqsBg}
\end{aligned}
\eea
where the densities are given by the equations
\bea
\begin{aligned}
\rho_{1a} & =\rho_a=\frac{1}{V}\sum_k\left\lbrace \frac{(X_5^2E_b+E_a\omega_1^2-E_aE_b^2)f(\omega_1)}{\sqrt{D}\omega_1}\right. 
 +\left.\frac{(X_5^2E_b+E_a\omega_2^2-E_aE_b^2)f(\omega_2)}{\sqrt{D}\omega_2} \right\rbrace \;, \\
\rho_{1b} & =\rho_b=\frac{1}{V}\sum_k\left\lbrace \frac{(X_5^2E_a+E_b\omega_1^2-E_a^2E_b)f(\omega_1)}{\sqrt{D}\omega_1}\right.
 +\left.\frac{(X_5^2E_a+E_b\omega_2^2-E_a^2E_b)f(\omega_2)}{\sqrt{D}\omega_2} \right\rbrace \;, \\
 \rho_{ab} & =\frac{2X_5}{V}\sum_k\left\lbrace \frac{(-X_5^2+E_aE_b+\omega_1^2)f(\omega_1)}{\sqrt{D}\omega_1}\right. 
 -\left.\frac{(-X_5^2+E_aE_b+\omega_2^2)f(\omega_2)}{\sqrt{D}\omega_2} \right\rbrace \;, \\
\eta & =\dsfrac{\rho_{ab}} {2\sqrt{\rho_a\rho_b}}, \quad \quad f(x)=1/(e^{\beta x}-1) \; .
\label{densBg}
\end{aligned}
\eea
Then the dispersion relations (\ref{dispX5X6}) reduce to the form
\bea
\begin{aligned}
&\omega_{1,2}=\sqrt{\frac{E_a^2+E_b^2}{2}+X_5^2\pm\frac{\sqrt{D}}{2}} \; ,\\
&D=(E_a+E_b)^2[4X_5^2+(E_a-E_b)^2] \; ,
\label{dispBg}
\end{aligned}
\eea
with
\bea
E_a=\veps_a(k)-\mu_{\rm eff}^{(a)}, \ \ \ \ \ \ E_a=\veps_b(k)-\mu_{\rm eff}^{(b)} \; .
\eea
This spectrum is real and positive, provided the expression under the square root does not
become negative. It is convenient to test the positiveness of the expression
$\omega_1^2 \omega_2^2$. Then from Eq. (\ref{dispBg}) one easily obtains the condition
\be
\omega_1^2 \omega_2^2=(E_aE_b-X_5^2)^2\geqslant 0 \; .
\ee
Since $\omega_1$ is positive, both $\omega_1^2$ and $\omega_2^2$ are positive simultaneously
for any $g_{ab}$ and temperature $T\geqslant T_c$; hence the spectra are real and positive.

Note that, when $g_{ab}=0$, Eqs. (\ref{densBg}) turn into the well-known
expressions
\bea
\begin{aligned}
&\rho_{1a}=\rho_a=\frac{1}{V}\sum_k\frac{1}{e^{\beta(\veps_a(k)-\mu_{\rm eff}^{(a)})}-1} \; ,\\
&\rho_{1b}=\rho_b=\frac{1}{V}\sum_k\frac{1}{e^{\beta(\veps_b(k)-\mu_{\rm eff}^{(b)})}-1} \; ,
\label{dens1comp}
\end{aligned}
\eea
where $\mu_{\rm eff}^{(a)}=\mu_a-2\rho_a g_a$.

The spectrum of collective excitations of a binary mixture of normal components has been
studied in the random-phase approximation in Ref. \ci{Yukalov_APPA}. In that approximation,
for the case of contact interactions, the dynamic condition for mixture stability is found
to coincide with the inequality $g^2_{ab} < g_a g_b$, being practically independent of
temperature. 

\subsection{Near critical temperature}

Let $T_c^{a}$ and $T_c^b$ be the BEC transition temperatures for the corresponding components.
For concreteness, we assume that $T_c^{a}\leqslant T_c^b$. In fact, $T_c^{a}$ corresponds
to the point where the effective chemical potential vanishes, $\mu_{\rm eff}^{(a)}=-X_1=0$.
From the stability condition $\Delta_a\Delta_b > \Delta_{ab}^2$, that is,
$X_5^2\leqslant X_1 X_3$, it is understood that the system can be stable only for 
 $X_5=0$, with the sound velocities
\bea
c_1^2=\frac{\Delta_b}{m_b}\geqslant 0, \ \ \ \ \ \ \
c_2^2=0.
\eea
This temperature can be evaluated from Eqs. (\ref{dens1comp}) as
\be
\rho_{1a}=\rho_a=\frac{1}{V}\sum_k\frac{1}{e^{\veps_a(k)/T_c^{a}}-1}.
\ee
From this equation it is seen that in the present approximation, like in many versions of
mean-field approximations \cite{andersen}, there is no shift of critical temperature due
to intercomponent coupling constant $g_{ab}$, i.e., $T_c(g_{ab})=T_c(g_{ab}=0)$.

For completeness, at the end of this section we present explicit expressions for the free
energy
$ F=\Omega+\mu N=\Omega+\mu_aN_a+\mu_bN_b $, which has the form
\bea
\begin{aligned}
F(T&<T_c)=F_0+F_{ZM}+F_T, \\
F_0=&\frac{V\rho_a^2g_a}{2}(1+n_{1a}^2-{\tilde m}_a^2-2n_{1a}{\tilde m}_{a})+\frac{V\rho_b^2g_b}{2}(1 
+n_{1b}^2-{\tilde m}
_b^2-2n_{1b}{\tilde m}_{b})+\frac{V}{2}\rho_a\rho_b(2-{\tilde m}_{ab}^2), \\
F_{ZM}&=\frac{1}{2}\sum_k\Bigg\{\omega_1+\omega_2-\veps_k-\Delta_a-\Delta_b 
  +\frac{\nu_2^2 \Delta_a^2+\nu_1^2\Delta_b^2+4\Delta_{ab}^2 \nu_1 \nu_2}{2\nu_1 \nu_2 \veps(k)} \Bigg\}, \\
F_T=&T\sum_k\left[\ln(1-e^{-\omega_1\beta})+\ln(1-e^{-\omega_2\beta})\right], \\
F(T&>T_c)=Vg_a\rho_a^2+Vg_b\rho_b^2+g_{ab} V \rho_a\rho_b+F_T,
\label{F6EN}
\end{aligned}
\eea
where $n_{1a}=\rho_{1a}/\rho_a$, ${\tilde m}_a=\sigma_a/\rho_a$,
${\tilde m}_{ab}=\sigma_{ab}/\rho_a$, and the dispersions $\omega_1$ and $\omega_2$ for the
BEC and normal phases are given in Eqs. (\ref{dispBEC}) and (\ref{dispBg}), respectively.
The explicit expressions for $F_{ZM}$, referred in the literature as zero mode energy,  can
be found, e.g., in Refs. \cite{Petrov15, larsen1963}. Note that the present approach
includes by itself not only the Lee-Huang-Yang (LHY) term  \ci{LHY}, but also the corrections
beyond the LHY approximation due to taking account of anomalous densities. Particularly,
the LHY term can be obtained expanding $F_{ZM}$ in powers of the coupling parameters.


Concluding the present section, let us summarize the conditions of stability for a two-
component uniform Bose system:\\

(i) At temperatures below the critical one, when the system is in the condensed phase, 
the mixture is stable, provided the general condition 
\be
\dsfrac{\Delta_{a}(\gamma, T) \Delta_{b}(\gamma, T)}{\Delta_{ab}^{2}(\gamma, T)} \ge 1
\lab{condgen}
\ee  
holds. Here  the self-energies $\Delta_{a}(\gamma, T)$, $\Delta_{b}(\gamma, T),$ and 
$\Delta_{ab}(\gamma, T)$ are the solutions to  Eqs. \re{Eqdelta}.

(ii) The inequality \re{condgen} may be replaced by 
\be
\dsfrac{g_{a}  g_{b} }{g_{ab}^2} \ge 1
\lab{cong12}
\ee  
for very dilute gases, where the Bogoliubov approximation is valid. \\
In the next section it will be shown that, for a balanced symmetric Bose mixture, the 
inequality \re{condgen} may be represented as an expansion in  powers of $\gamma$. 


\section{Balanced Symmetric Bose mixtures}

The case of a binary superfluid gas with two symmetric components consisting of $^{23}$Na
in an equal mixture of two hyperfine ground states has been recently realized experimentally
by Kim \textit{et al.}  \cite{kim20}. So we assume that $g_a=g_b=g$, $g_{ab}=\overline{g}_{ab}g_a$,
$m_a=m_b=m$, $\veps_a(k)=\veps_b(k)=\veps(k)=k^2/2m$, and $\rho_a=\rho_b=\rho/2$, where $N=\rho V$
is the total number of atoms in the mixture. Note that, treating the anomalous averages, we
resort to the standard way of regularization by employing the method of counterterms that is
equivalent to the dimensional regularization \cite{andersen, yukechaya}.


\subsection{Zero temperature}

At zero temperature, the densities are simplified to
\nopagebreak
\begin{widetext}
\bea
\begin{aligned}
 n_{1a}=& \frac{\rho_{1a}(T=0)}{\rho_a}=\frac{\rho_{1b}(T=0)}{\rho_a}=\frac{1}{2V\rho_a}\sum_k\left\lbrace \frac{\Delta_a+\veps(k)+\Delta_{ab}}{2\omega_1}+
\frac{\Delta_a+\veps(k)-\Delta_{ab}}{2\omega_2}-1 \right\rbrace \\
=&\frac{m^3(c_1^3+c_2^3)}{6\pi^2\rho_a}=n_{1b} \; , \\
 {\tilde m}_a=&\frac{\sigma_a(T=0)}{\rho_a}=\frac{\sigma_b(T=0)}{\rho_b}=-\frac{1}{2V\rho_a}\sum_k\left\lbrace \frac{\Delta_a+\Delta_{ab}}{2\omega_1}+\frac{\Delta_a-\Delta_{ab}}{2\omega_2}-\frac{\Delta_a}{\veps(k)} \right\rbrace \\
=&3n_{1a}(T=0) \; ,\\
n_{ab}=&\frac{\rho_{ab}(T=0)}{\sqrt{\rho_a\rho_b}}=\frac{1}{2V\rho_a}\sum_k\left\lbrace \frac{\veps(k)+\Delta_a+\Delta_{ab}}{\omega_1}-\frac{\veps(k)+\Delta_a-\Delta_{ab}}{\omega_2} \right\rbrace=\frac{m^3(c_1^3-c_2^3)}{3\pi^2\rho_a} \; ,\\
 {\tilde m}_{ab}=&\frac{\sigma_{ab}(T=0)}{\sqrt{\rho_a\rho_b}}=n_{ab} \; ,
\label{denseT}
\end{aligned}
\eea
\end{widetext}
where we have introduced the sound velocities
\begin{equation}
c^2_{1,2}=(\Delta_a \pm \Delta_{ab})/m \; ,
\label{soundspeed}
\end{equation}
which satisfy the following equations, derived from Eqs. (\ref{DeltaSimp}) and (\ref{soundBEC}):
\bea
\begin{aligned}
&c_1^3-c_2^3(\overline{g}_{ab}-1)-\frac{3\pi^2c_1^2}{gm^2}+\frac{3\pi^2\rho_a(\overline{g}_{ab}+1)}{m^3}=0  \; , \\
&c_1^3+c_2^3(\overline{g}_{ab}+1)-\frac{3\pi^2c_2^2}{gm^2}-\frac{3\pi^2\rho_a(\overline{g}_{ab}-1)}{m^3}=0 \; .
\label{C12T0}
\end{aligned}
\eea
These equations can be rewritten in the dimensionless form as
\bea
\begin{aligned}
&s_1^3+(1-\overline{g}_{ab})s_2^3-\frac{3\pi s_1^2}{4}+\frac{3\pi^2\gamma (\overline{g}_{ab}+1)}{2}=0 \; , \\
&s_1^3+(1+\overline{g}_{ab})s_2^3-\frac{3\pi s_2^2}{4}-\frac{3\pi^2\gamma (\overline{g}_{ab}-1)}{2}=0 \; ,
\label{S12T0}
\end{aligned}
\eea
where $\gamma=\rho a_s^3$, $\rho=2\rho_a$ is the total density of the whole binary system,
$a_s=mg/4\pi$, and $s_{1,2}=c_{1,2} ma_s$. From the Bogoliubov approximation, it is
known that the system becomes unstable for $\overline{g}_{ab}>1$. Here we study the
problem of possible corrections to this criterion due to quantum fluctuations.

\begin{figure}[h]
\begin{minipage}[h]{0.48\linewidth}
  \includegraphics[clip,width=\columnwidth]{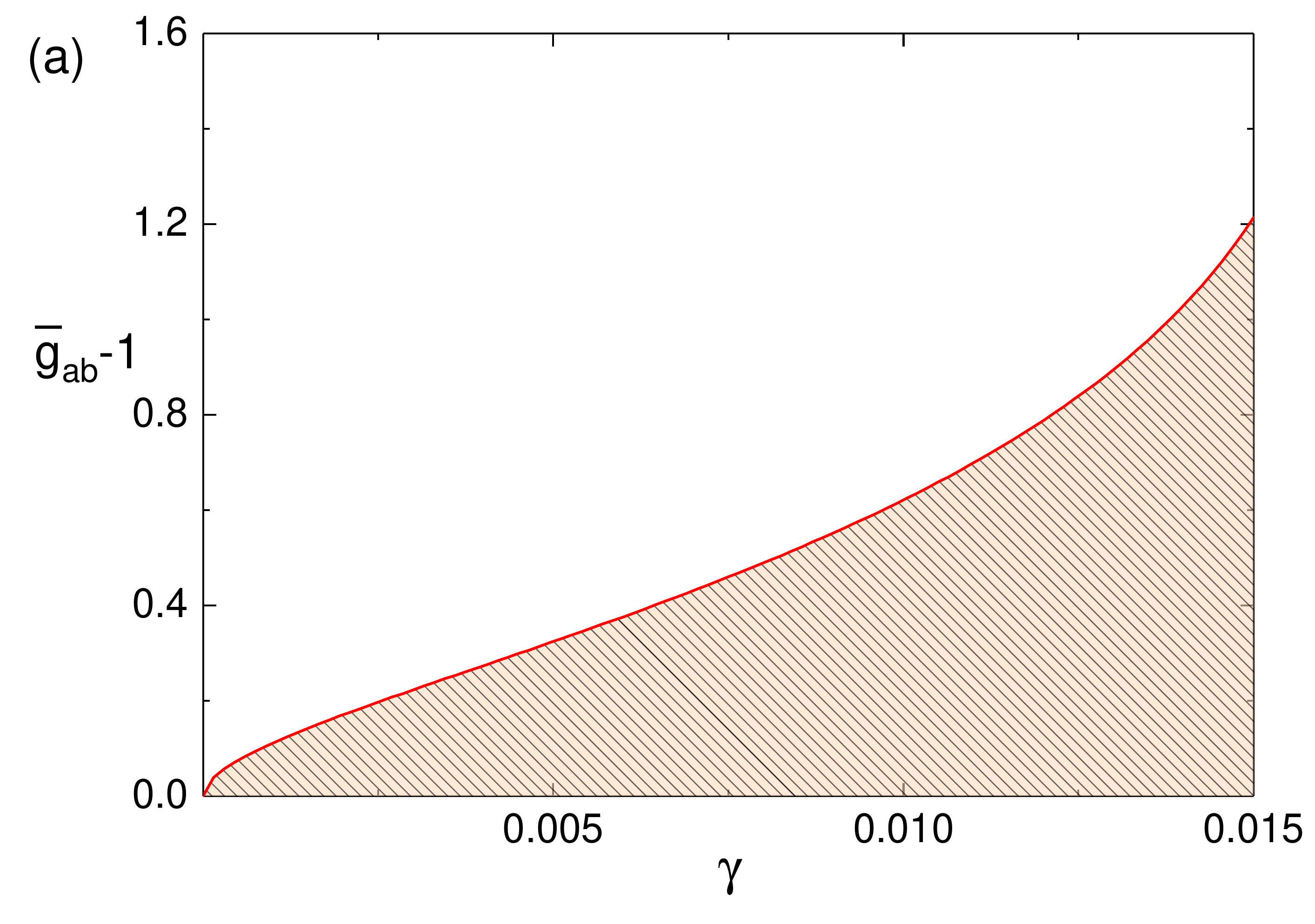}%
\\
\end{minipage}
\hfill
\begin{minipage}[h]{0.48\linewidth}
  \includegraphics[clip,width=\columnwidth]{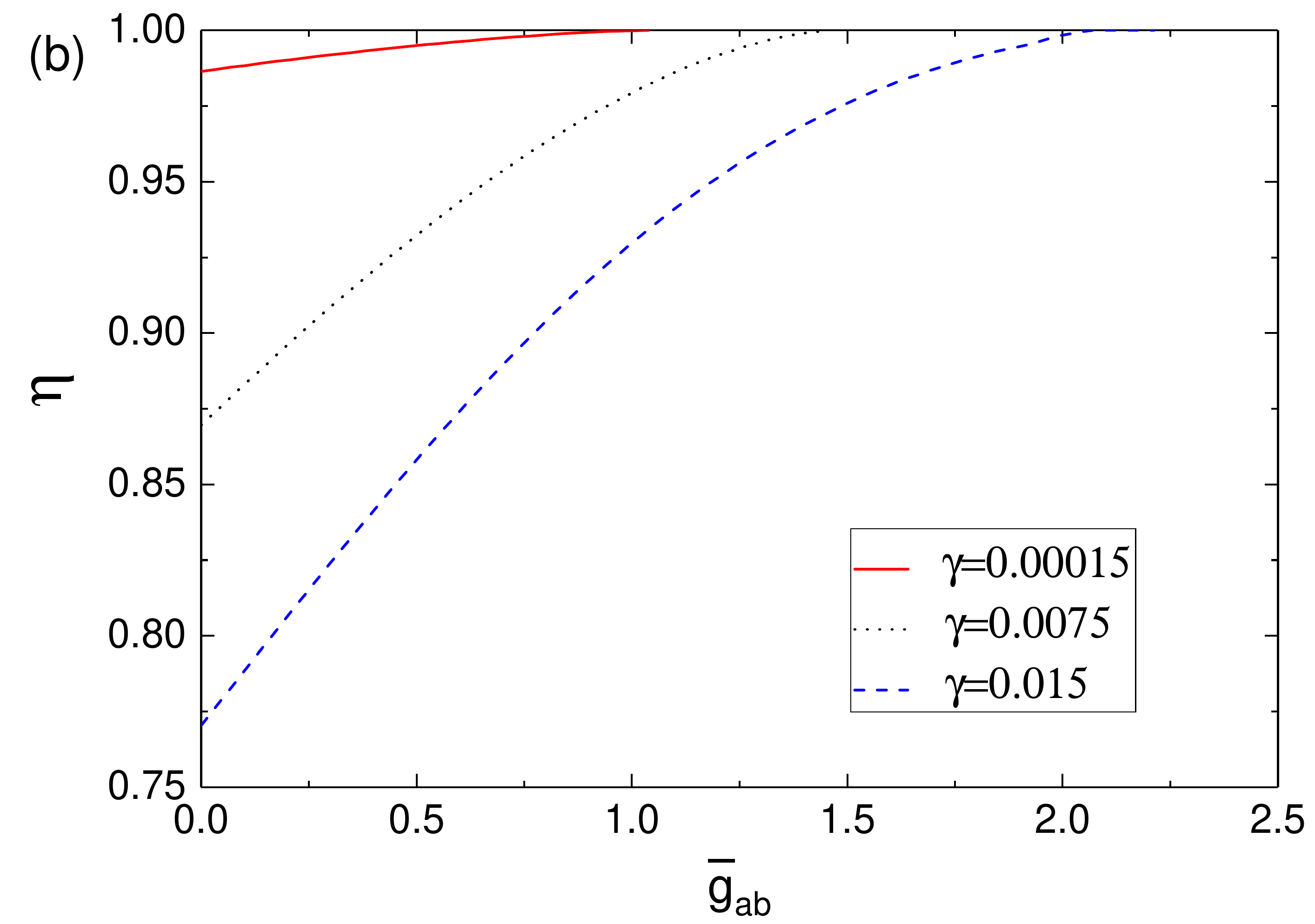}%
\\
\end{minipage}
\hfill

\caption{(a)  The phase diagram of a symmetric binary Bose system with repulsive interactions at
zero temperature. The shaded region corresponds to the stable, miscible phase. \\
 (b)  Overlap parameter $\eta$ vs $\overline{g}_{ab}=g_{ab}/g$ for three different values
of the gas parameter: $\gamma=0.15\times 10^{-3}$ (solid line), $\gamma=0.75\times 10^{-2}$
(dotted line), and $\gamma=0.15\times 10^{-1}$ (dashed line).
}
  \label{Fig1}
\end{figure}

To find the answer to this question one has to consider the dispersions:
\bea
\begin{aligned}
&\omega_1^2=\veps(k)[\veps(k)+2(\Delta_a+\Delta_{ab})] \; , \\
&\omega_2^2=\veps(k)[\veps(k)+2(\Delta_a-\Delta_{ab})] \; .
\label{dispT0}
\end{aligned}
\eea
It is seen that the boundary of stability is defined by the condition $\Delta_a=\Delta_{ab}$,
since for $\Delta_a <\Delta_{ab}$ the sound velocity $c_2$ becomes negative. This is also
seen from the general condition $\Delta_a\Delta_b\leqslant\Delta_{ab}^2$, with
$\Delta_a=\Delta_b$. In other words, the boundary of stability in the phase diagram
$(\overline{g}_{ab},\gamma)$ lies on the line $c_2=0$, i.e.,
$s_2(\overline{g}_{ab}^{*},\gamma^{*})\equiv 0$. From Eqs. (\ref{S12T0}), one obtains
\bea
\begin{aligned}
 & (s_1^{*})^3- \frac{3\pi}{8}(s_1^{*})^2+\frac{3\pi^2\gamma^{*}}{2}=0 \; , \\
& \overline{g}_{ab}^{*}=\frac{(s_1^{*})^2}{4\pi\gamma^{*}} \; .
\label{alfastar}
\end{aligned}
\eea
where the asterisk indicates the threshold values of the parameters, corresponding to the
boundary of stability for the symmetric binary mixture. In Fig. \ref{Fig1}(a), we present
the phase diagram on the $(\overline{g}_{ab},\gamma)$ plane (solid line). It is seen that,
due to quantum fluctuations, the system at $T=0$ remains stable even, for example, at
$\overline{g}_{ab}^*(\gamma\approx 0.013)\approx 1.9$. This is one of the main results of
the present work.

For small $\gamma$, one may use the following expansion for $\overline{g}_{ab}^{*}$:
\be
\overline{g}_{ab}^{*}=1+\frac{16\sqrt{\gamma}}{3\sqrt{\pi}}+\frac{128\gamma}{3\pi}+\frac{3584\gamma^{3/2}}{9\pi^{3/2}}+O(\gamma^{5/2}) \; ,
\ee
in order to obtain the stability condition in the form
\be
g_{ab} \le g_a \left[ 1+\frac
{16\sqrt{\gamma}}
{3\sqrt{\pi}}+O(\gamma)\right] \; ,
\ee
which is valid for $\gamma \leq 0.005$.


The overlap parameter for the symmetric case at $T=0$ has the form
\be
\eta=n_{0a}+\frac{n_{ab}}{2}=1-n_{1a}+\frac{n_{ab}}{2}=1-\frac{2s_2^3}{3\pi^2\gamma} \; ,
\label{etaT0}
\ee
where we use Eqs. (\ref{denseT}) and (\ref{soundspeed}). On the boundary of stability,
$s_2=s_2^{*}=0$, and $\eta$ reaches its maximum value $\eta=1$ [see Fig. \ref{Fig1}(b)].

\begin{figure}[h]
\begin{minipage}[h]{0.49\linewidth}
  \includegraphics[clip,width=\columnwidth]{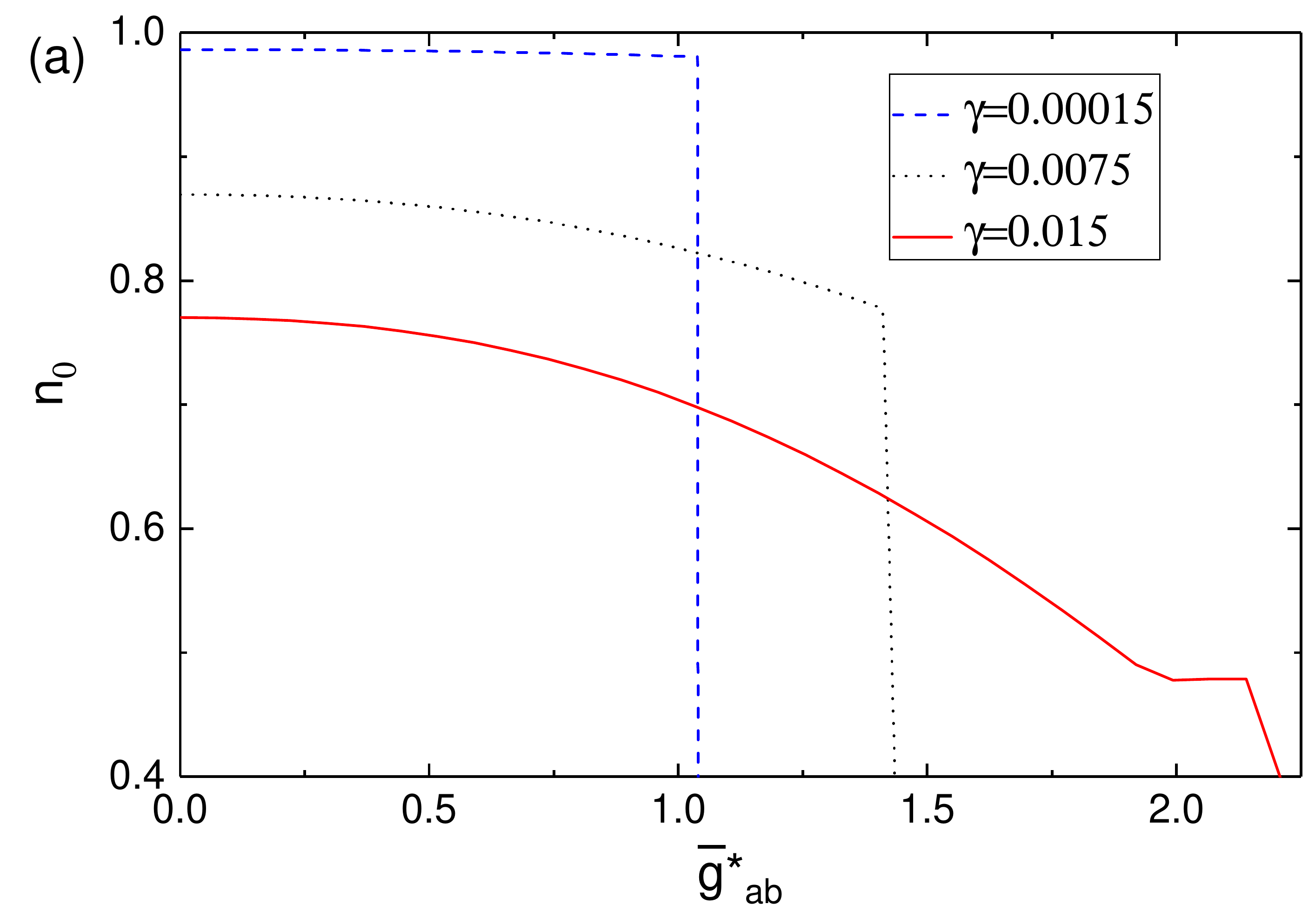}%
\\
\end{minipage}
\hfill
\begin{minipage}[h]{0.49\linewidth}
  \includegraphics[clip,width=\columnwidth]{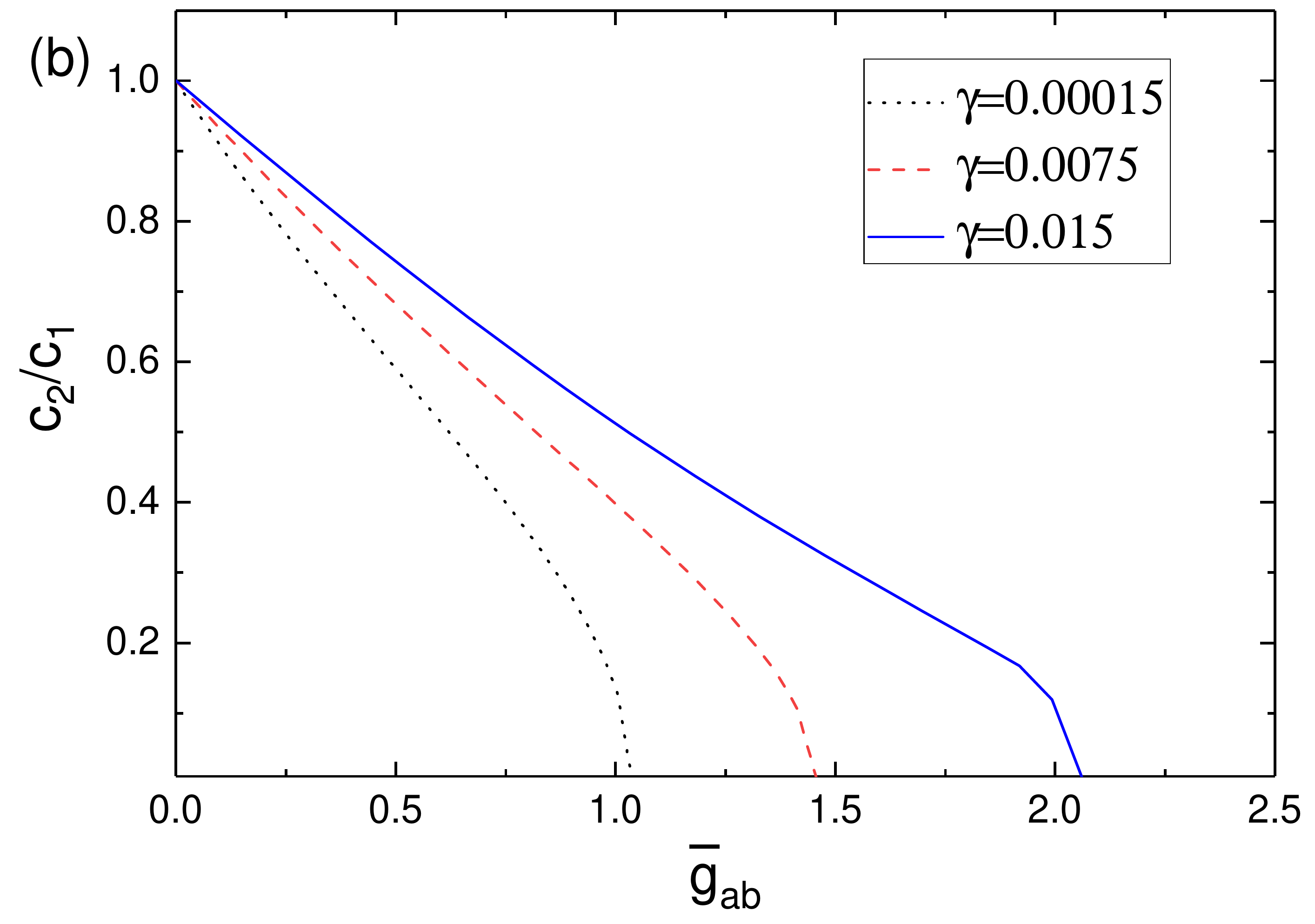}%
\\
\end{minipage}
\hfill

\caption{a)  The condensed fraction [$n_0=\rho_0/(\rho/2)$] at zero temperature vs $\overline{g}_{ab}=g_{ab}/g$ in the
interval $0 < \overline{g}_{ab} < \overline{g}_{ab}^{*}(\gamma)$ for different values of
$\gamma$.
(b)  The relative sound velocity $c_2/c_1$ vs $\overline{g}_{ab}$ for different values
of $\gamma$ at $T=0$.}
  \label{Fig2}
\end{figure}

Close to the phase transition point, the condensed fraction can be presented as
\be
\left. n^*_{0}(T=0)=1-\frac{\rho_{1a}}{\rho_a}\right|_{\overline{g}_{ab}\rightarrow \overline{g}_{ab}^*}=1-\frac{8\sqrt{\gamma}}{3\sqrt{\pi}}-\frac{64\gamma}{3\pi}+O(\gamma^{3/2}) \; .
\ee
In Fig. \ref{Fig2}(a), we present the condensed fraction vs $\overline{g}_{ab}$. It is seen
that intercomponent repulsion $g_{ab}$ tends to destroy BEC, repelling the condensed particles.

The present work would not be complete without a comparison with the experiment performed by
Kim \textit{et al.} \cite{kim20}. The authors studied the mixture of atoms with two hyperfine ground
states of $^{23}$Na and measured the sound velocities $c_1=3.23 $ mm/s and $c_2=0.70  $ mm/s
by fixing the relative coupling constant $\overline{g}_{ab}=0.93$ and the gas parameter
$\gamma\approx 1.4\times 10^{-6}$. For this set of parameters from Eqs. (\ref{C12T0}), we get
the following values for the sound velocities: $c_1=3.91 \ $mm/s, $c_2=0.75 \ $mm/s, which
are rather close to the experimental data. To make further prediction, we have calculated
the relative sound velocity $c_2/c_1$ vs $\overline{g}_{ab}$ for three different values of
$\gamma$. The results are presented in Fig. \ref{Fig2}(b). It is seen that $c_2/c_1$ reduces
with increasing $\overline{g}_{ab}$ and vanishes at $\overline{g}_{ab}=\overline{g}_{ab}^*$,
where the phase separation occurs.

\subsection{Finite temperature BEC in a balanced symmetric binary mixture}

Setting $m_a=m_b=m$, $\rho_a=\rho_b=\rho/2$, $g_a=g_b=g$, and $g_{ab}=\overline{g}_{ab}g$
in Eqs. (A11) -- (A16), we can obtain the following expressions for the densities at finite
temperatures,

\bea
 n_{1a}=&&\frac{\rho_{1a}}{\rho_a}=\frac{m^3(c_1^3+c_2^3)}{6\pi^2\rho_a}+\frac{1}{2V\rho_a}\sum_k \left[\frac{mc_1^2+\veps(k)}{\omega_1}f(\omega_1)\right. \nonumber\\
 &&+\left.\frac{mc_2^2+\veps(k)}{\omega_2}f(\omega_2) \right] \; , \nonumber\\
 {\tilde m}_a=&&\frac{\sigma_a}{\rho_a}=\frac{m^3(c_1^3+c_2^3)}{2\pi^2\rho_a}-\frac{m}{2V\rho_a}\sum_k[c_1^2f(\omega_1)+c_2^2f(\omega_2)] \; ,\nonumber\\
 n_{ab}=&&\frac{\rho_{ab}}{\rho_{a}}=\frac{m^3(c_1^3-c_2^3)}{3\pi^2\rho_a}+\frac{1}{V\rho_a}\sum_k\left[\frac{c_1^2m+\veps(k)}{\omega_1}f(\omega_1)\right. \nonumber\\
 &&+\left.\frac{c_2^2m+\veps(k)}{\omega_2}f(\omega_2)\right] \; ,
\label{densTf}
\eea

as well as for the overlap parameter $\eta$,
\be
\eta=1-\frac{m^3c_2^3}{3\pi^2\rho_a}-\frac{1}{V\rho_a}\sum_k\frac{mc_2^2+\veps(k)}{\omega_2}f(\omega_2) \; ,
\label{etaTf}
\ee
where the dispersions, given in terms of the sound velocities, are
\be
\omega_{1,2}=\sqrt{\veps(k)(\veps(k)+2mc_{1,2}^2)} \; .
\ee
Note that close to the critical temperature, where $c_2=0$ and $\omega_2=\veps(k)$, the
overlap parameter $\eta$ tends to zero:
\be
\eta(T\rightarrow T_c)=1-\frac{1}{\rho_aV}\sum_k\frac{1}{e^{\veps(k)/T_c}-1}=1-1=0 \; ,
\ee
as is shown for the general case in Sec. III.
As to the  main equations (\ref{Eqdelta}) they are simplified as
\bea
\begin{aligned}
& \Delta_{a}=\Delta_{b}=\frac{m}{2}(c_1^2+c_2^2)=g\rho_a[1-n_{1a}+{\tilde m}_a] \; ,\\
& \Delta_{ab}=\frac{m}{2}(c_1^2-c_2^2)=g\overline{g}_{ab}\rho_a[1-n_{1a}+\frac{n_{ab}}{2}] \; .
\label{EqDelTf}
\end{aligned}
\eea

For practical calculations, it is convenient to solve the system of the following
dimensionless equations:
\bea
\begin{aligned}
& s_1^3+s_2^3-\frac{3\pi}{8}(s_1^2+s_2^2)+\frac{3\pi^2\gamma}{2}-\frac{3\pi^2a_s}{2mV} \\
&\times\sum_k\left[\frac{a_s^2m\veps(k)+2s_1^2}{\omega_1}f(\omega_1)\right.\\&+\left.\frac{a_s^2m\veps(k)+2s_2^2}{\omega_2}f(\omega_2) \right]=0 \; , \\
& s_1^2-s_2^2+\frac{8\overline{g}_{ab}s_2^3}{3\pi}-4\pi\overline{g}_{ab}\gamma+\frac{8\pi\overline{g}_{ab}a_s}{mV} \\
&\times\sum_k\frac{s_2^2+a_s^2m\veps(k)}{\omega_2}f(\omega_2)=0 \; ,\end{aligned}
\lab{vels}
\eea
where $\omega_{1,2}=\veps^2(k)+2\veps(k)s_{1,2}^2/{ma_s^2}$, with respect to the dimensionless
sound velocities $s_i=c_ima_s$ and then evaluate the densities from Eqs. (\ref{densTf}) and
(\ref{etaTf}).

In the previous section we studied the boundary of stability [Fig.\ref{Fig1}(a)] at zero
temperature. As is clear, at finite temperature the stability condition
$\Delta_a(\overline{g}_{ab},\gamma, T)\leqslant \Delta_{ab}(\overline{g}_{ab}, \gamma, T)$
can also be violated. Now the stability becomes lost at a certain point with
$\overline{g}_{ab}=\overline{g}_{ab}^*$ for a given gas parameter $\gamma$ and temperature $T$.
At this point, according to Eqs. (\ref{EqDelTf}), we have $c_2=0$ and hence
$\overline{g}_{ab}^*$ satisfies the following equation:

\begin{equation}
  \overline{g}_{ab}^*=\frac{v_1^2}{2(1-t^{3/2})} \; ,
  \label{18.1}
\end{equation}
where we use the identity
\begin{equation}
\frac{1}{V}\sum_k f(\omega_2)=\frac{1}{V}\sum_k\frac{1}{e^{\veps(k)\beta}-1}=\rho_a t^{3/2} \; ,
\label{18.2}
\end{equation}
with $t=T/T_c$ and $T_c=2\rho_a^{2/3}\pi/m\eta(3/2)^{2/3}\approx 2.08\gamma^{2/3}/a_s^2m$.
Along the line of the stability boundary, the reduced sound velocity in Eq. (\ref{18.1}) is
$v_1=ma_sc_1/\sqrt{2\pi\gamma}$, and the densities are given by the following equations:
\bea
1-\frac{v_1^2}{2}+\frac{4v_1^3}{3}\sqrt{\frac{2\gamma}{\pi}}-\frac{t^{3/2}}{2}
-\frac{a_s}{Vm\gamma}\sum_k\frac{(a_s^2\veps(k)m+4\pi v_1^2\gamma)}{\omega_1 }f(\omega_1)=0 \; , \lab{eq82}
\\
n_{1a}^*=\frac{t^{3/2}}{2}+\frac{2}{3}\sqrt{\frac{2\gamma}{\pi}}v_1^3
 +\frac{a_s}{\pi \gamma V}\sum_k\frac{(2\pi\gamma v_1^2+\veps(k) ma_s^2)}{\omega_1 }f(\omega_1)=0 \; , 
\eea
\bea
{\tilde m}_a^*&=&\frac{2v_1^3\sqrt{2\gamma}}{\sqrt{\pi}}-\frac{2a_s\pi v_1^2}{mV}\sum_k\frac{1}{ \omega_1 }f(\omega_1) \; , \\
n_{ab}^*&=&-t^{3/2}+\frac{4v_1^3}{3}\sqrt{\frac{2\gamma}{\pi}}
+\frac{2a_s}{m\gamma V}\sum_k\frac{(2\pi\gamma v_1^2+\veps(k) ma_s^2)}{\omega_1 }f(\omega_1) \; .
\lab{eqv1}
\eea
The equation for $\eta^*$ is simplified as
\be
\eta^*=\eta(c_2=0)=1-t^{3/2}\neq 0 \; ,
\ee
where we use Eqs. (\ref{etaTf}) and (\ref{18.2}).

In Fig. 3, we present the phase diagram of a symmetric two-component BEC on the
($\overline{g}_{ab}, t$) plane for four values of $\gamma$. It is seen that,  instability
can occur at any temperature below the critical one depending on $g_{ab}=\overline{g}_{ab}g$.
For example, at $t=0.5$ the system remains in a miscible, stable phase before
$\overline{g}_{ab}$ reaches the value $\overline{g}_{ab}^*=1.625$ for $\gamma=0.01$.
Increasing further $\overline{g}_{ab}$ at this temperature leads to the phase transition to
an immiscible, but stable phase which has lower energy. It is seen from Fig. 3(a) that, for small
$\gamma$ the threshold value of $\overline{g}_{ab}$ is close to unity in agreement with
the Bogoliubov prediction, and it increases with increasing $\gamma$ due to the quantum
corrections. Below $\overline{g}_{ab}<\overline{g}_{ab}^*$, the system is miscible and
stable.

In Fig. 4, we plot the overlap parameter $\eta(t)$ for different values of
$\overline{g}_{ab}$ and $\gamma$. As is seen, $\eta$ is close to unity near zero temperature,
and rapidly decreases by increasing the temperature, to vanish at $T=T_c$.
 \begin{figure}[h]
\begin{minipage}[h]{0.48\linewidth}
\center{\includegraphics[width=1\linewidth]{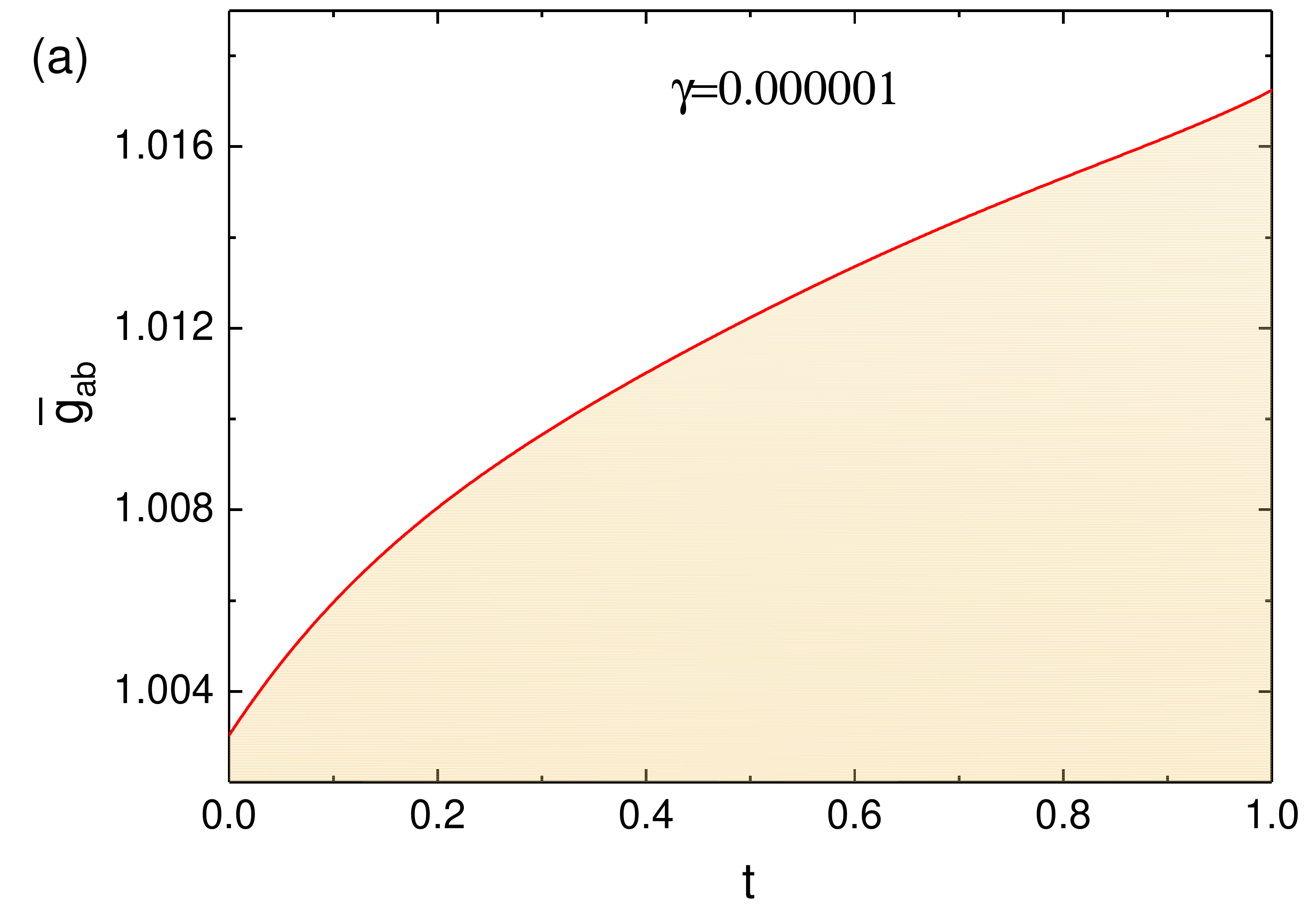}}
\\
\end{minipage}
\hfill
\begin{minipage}[h]{0.48\linewidth}
\center{\includegraphics[width=1\linewidth]{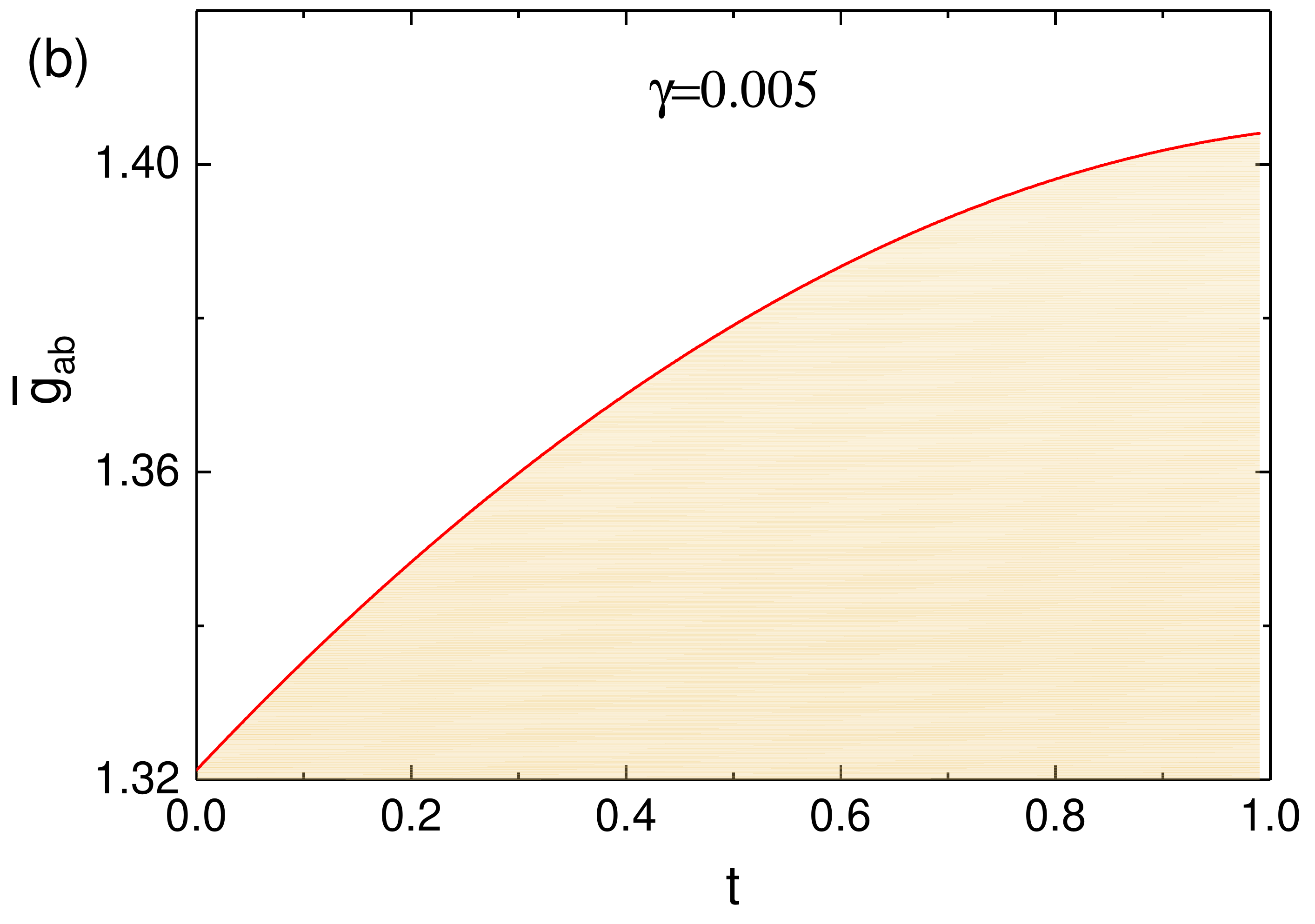}  }
\\
\end{minipage}
\hfill
\begin{minipage}[h]{0.48\linewidth}
\center{\includegraphics[width=1\linewidth]{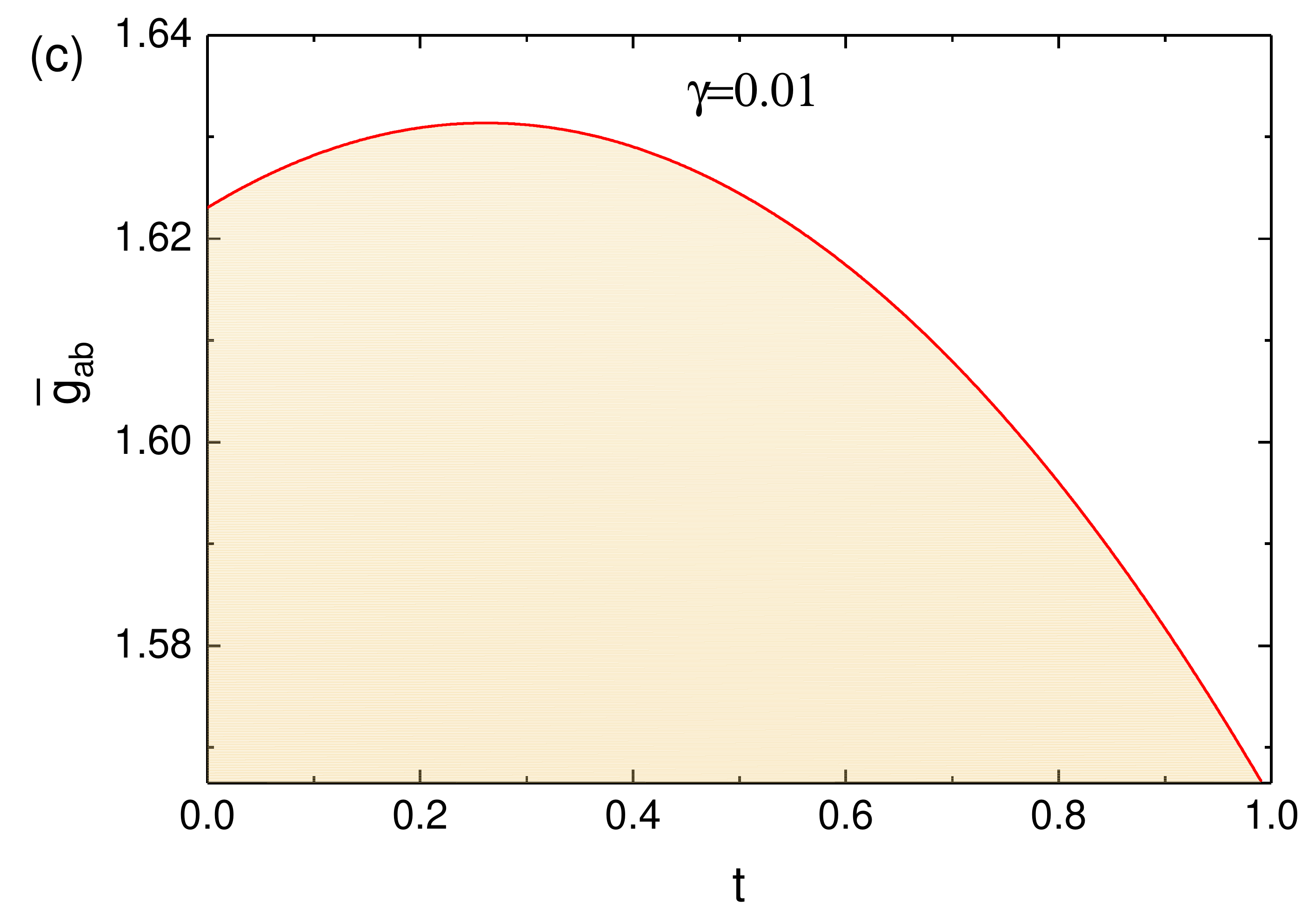}}
\\
\end{minipage}
\hfill
\begin{minipage}[h]{0.48\linewidth}
\center{\includegraphics[width=1\linewidth]{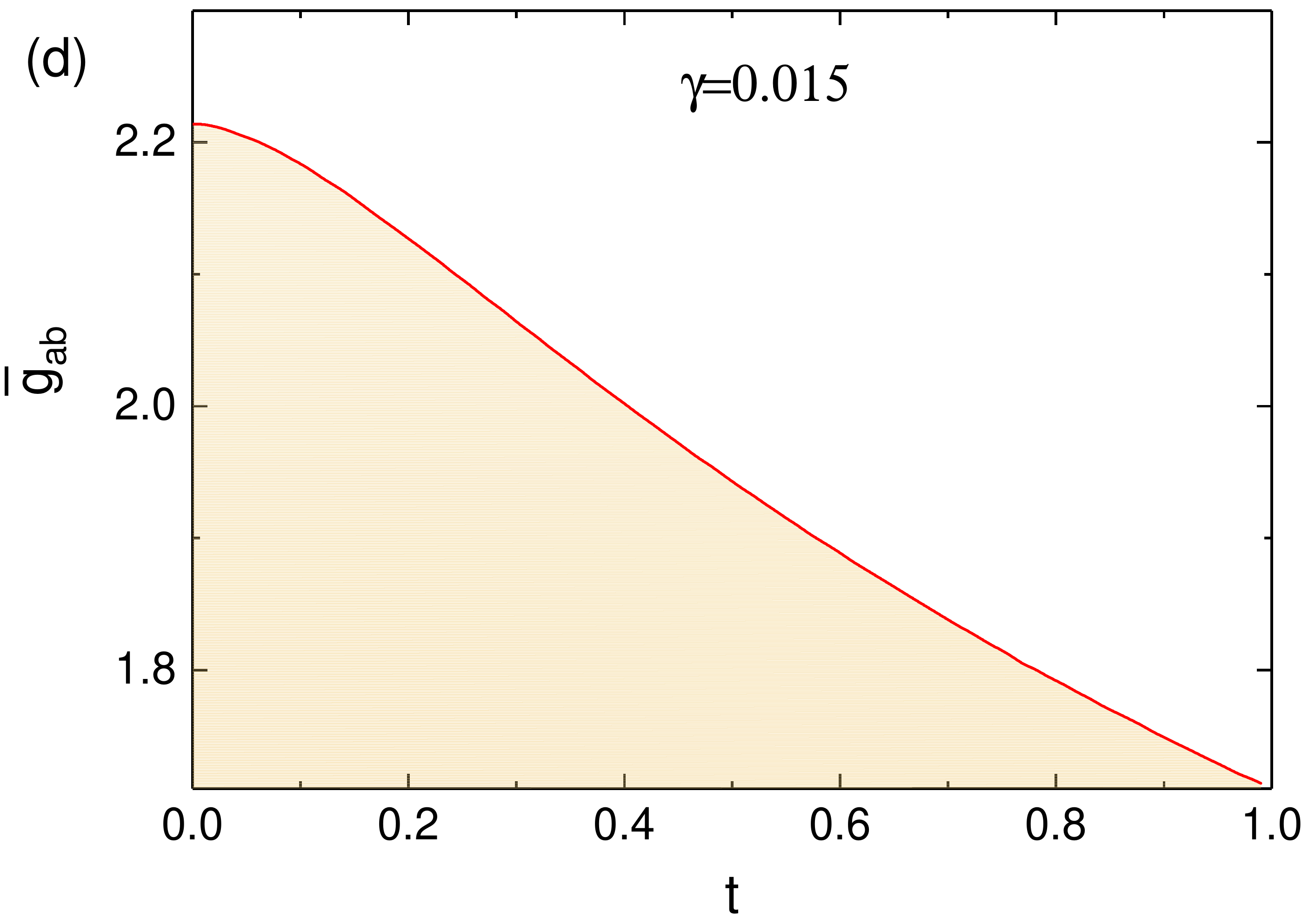}  }
\\
\end{minipage}
\hfill
\caption
{ Phase diagram of the balanced symmetric two-component Bose mixture
on the $(\overline{g}_{ab}=g_{ab}/g,   t=T/T_c)$ plane for different gas parameters: (a) $\gamma=0.000001$,
(b) $\gamma=0.005$, (c) $\gamma=0.01$,  and (d) $\gamma=0.015$. The shaded region corresponds
to a stable miscible state.
}
  \label{Fig3}
\end{figure}
\begin{figure}[h]
\begin{minipage}[h]{0.48\linewidth}
\center{\includegraphics[width=1\linewidth]{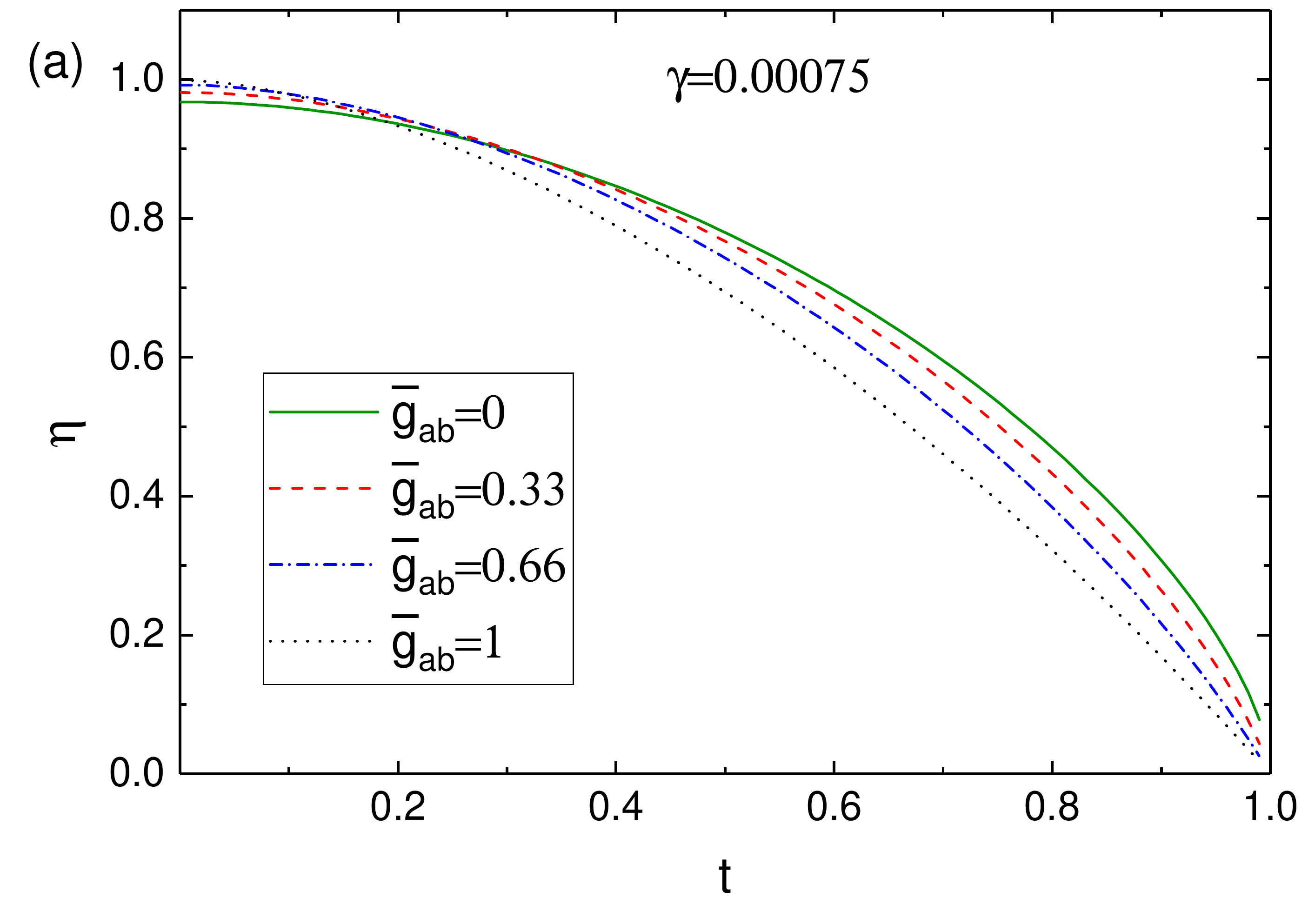} }
\\
\end{minipage}
\hfill
\begin{minipage}[h]{0.48\linewidth}
\center{\includegraphics[width=1\linewidth]{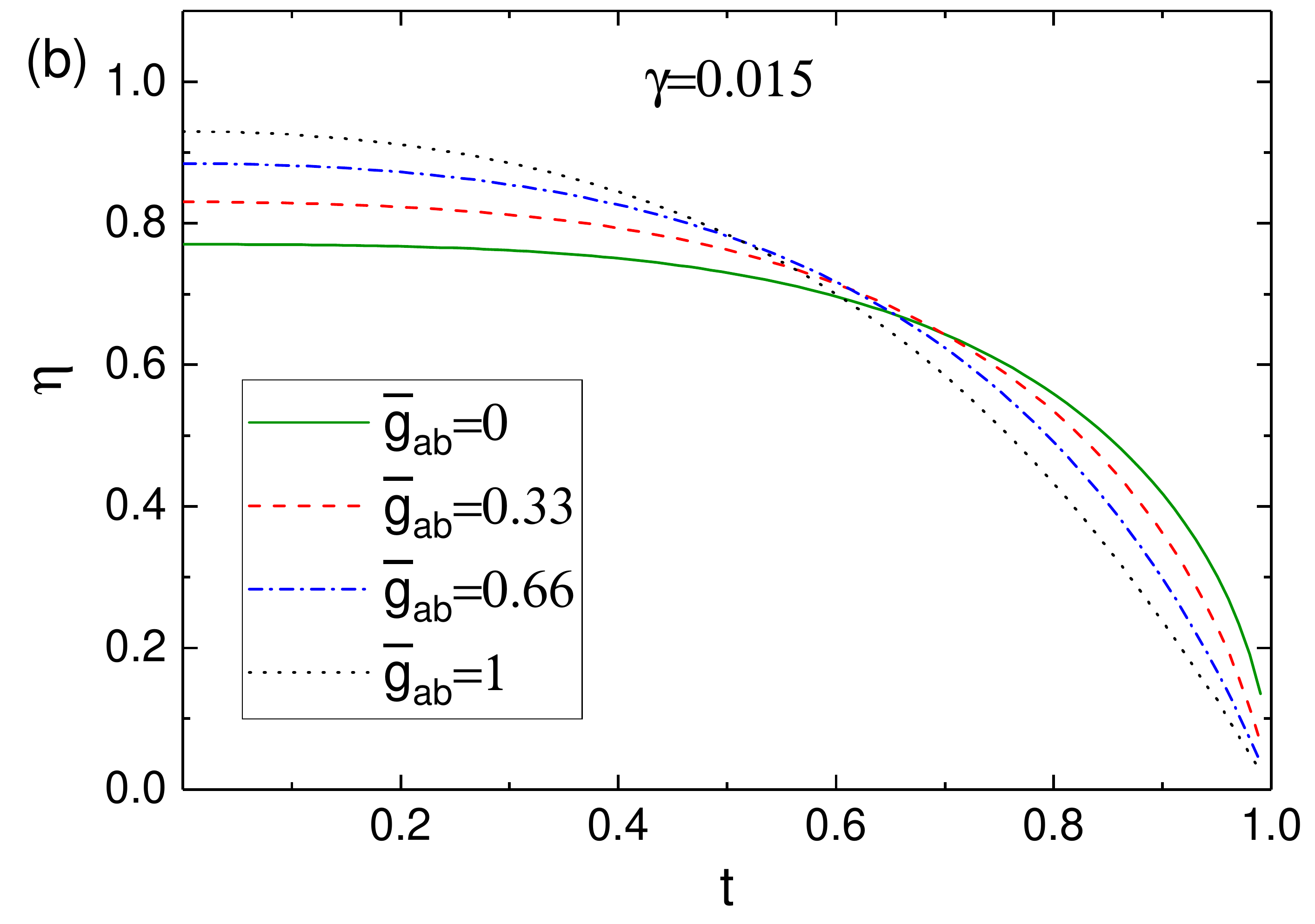}  }
\\
\end{minipage}
\hfill

\caption{\label{Fig4}
The overlap parameter vs temperature for (a) $\gamma=7.5\times 10^{-3}$  and (b) $\gamma=0.015$.
}
\end{figure}
 \begin{figure}[h]
\begin{minipage}[h]{0.47\linewidth}
\center{\includegraphics[width=1\linewidth]{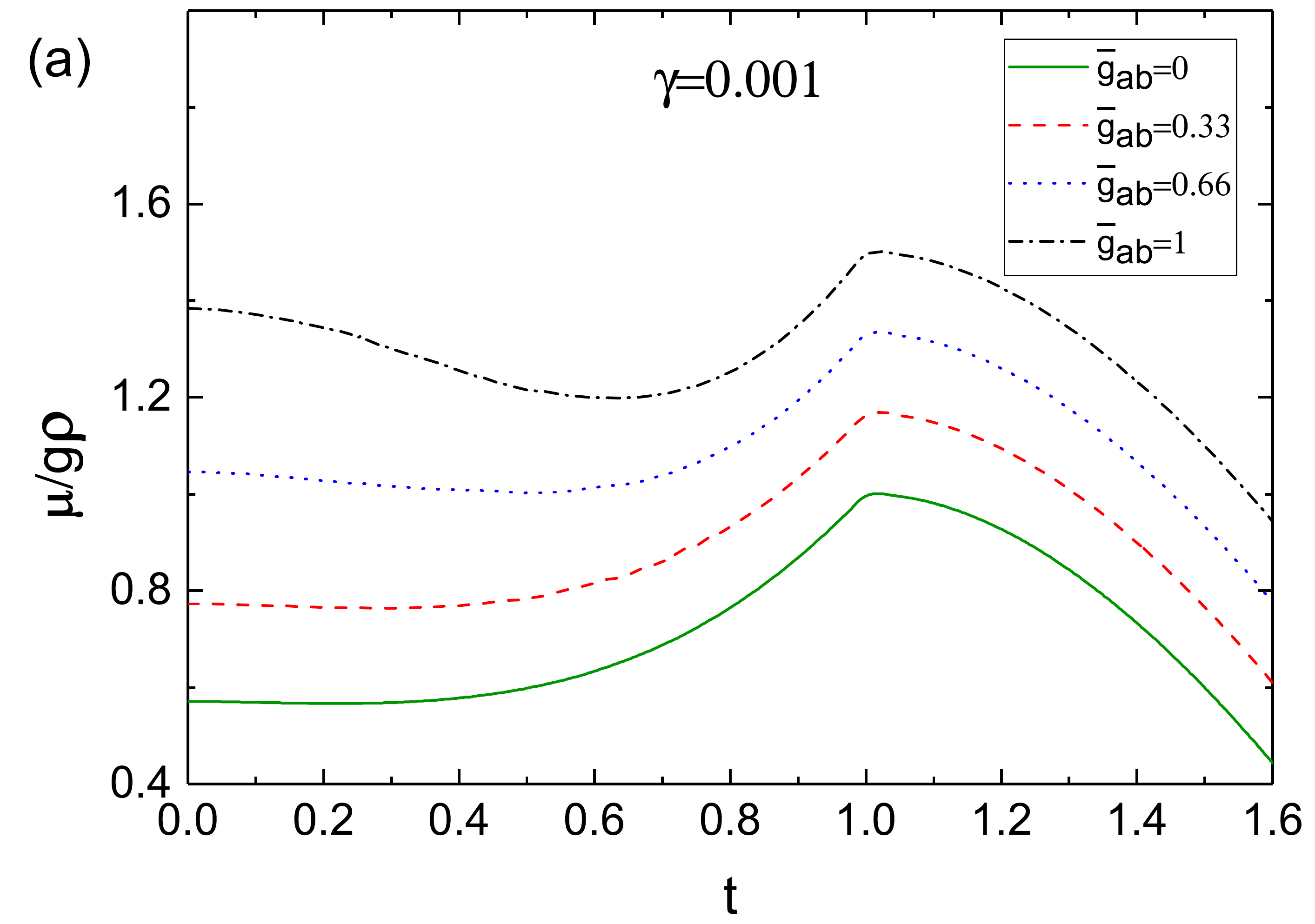} }
\\
\end{minipage}
\hfill
\begin{minipage}[h]{0.47\linewidth}
\center{\includegraphics[width=1\linewidth]{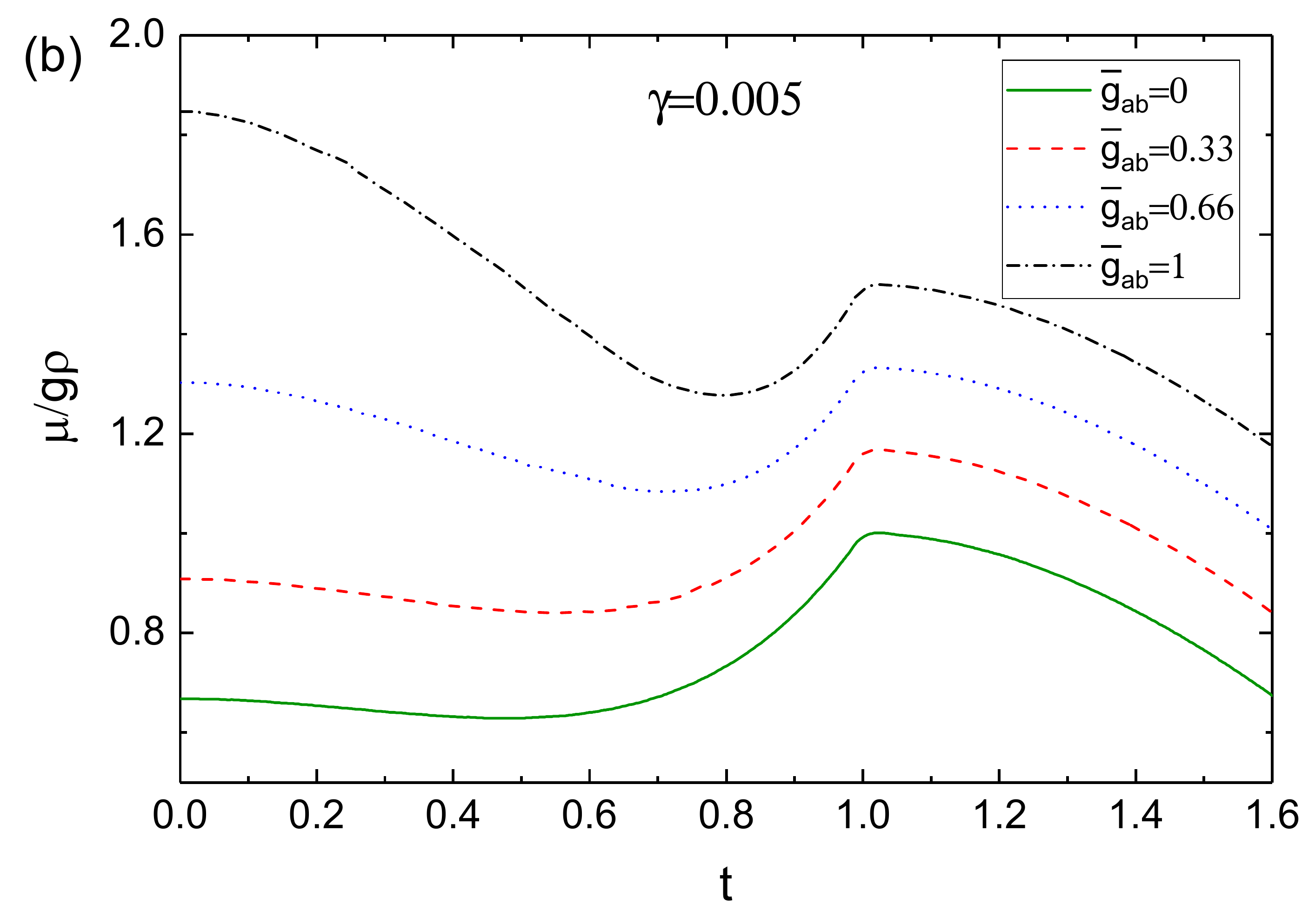}  }
\\
\end{minipage}
\hfill
\begin{minipage}[h]{0.47\linewidth}
\center{\includegraphics[width=1\linewidth]{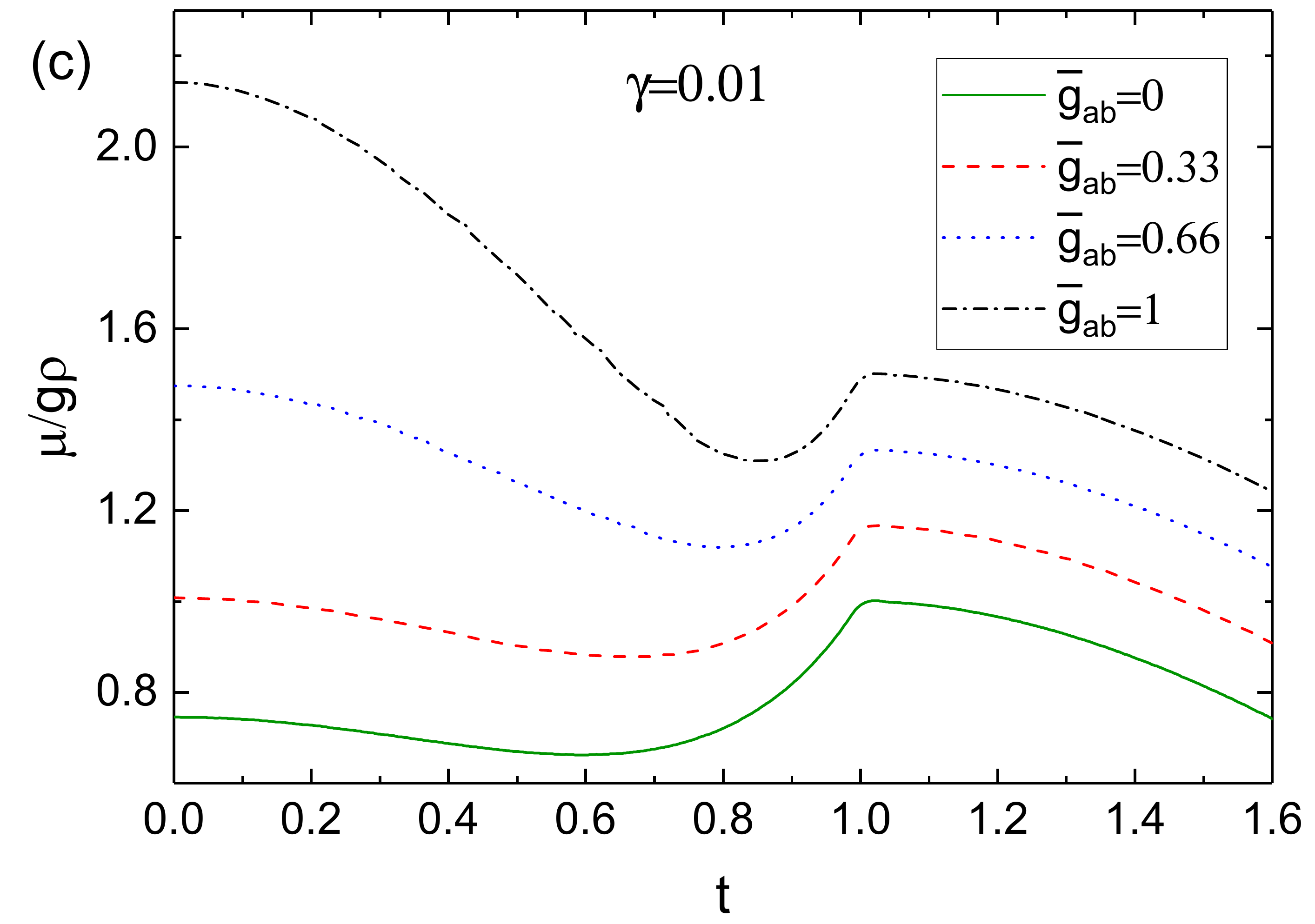}}
\\
\end{minipage}
\hfill
\begin{minipage}[h]{0.47\linewidth}
\center{\includegraphics[width=1\linewidth]{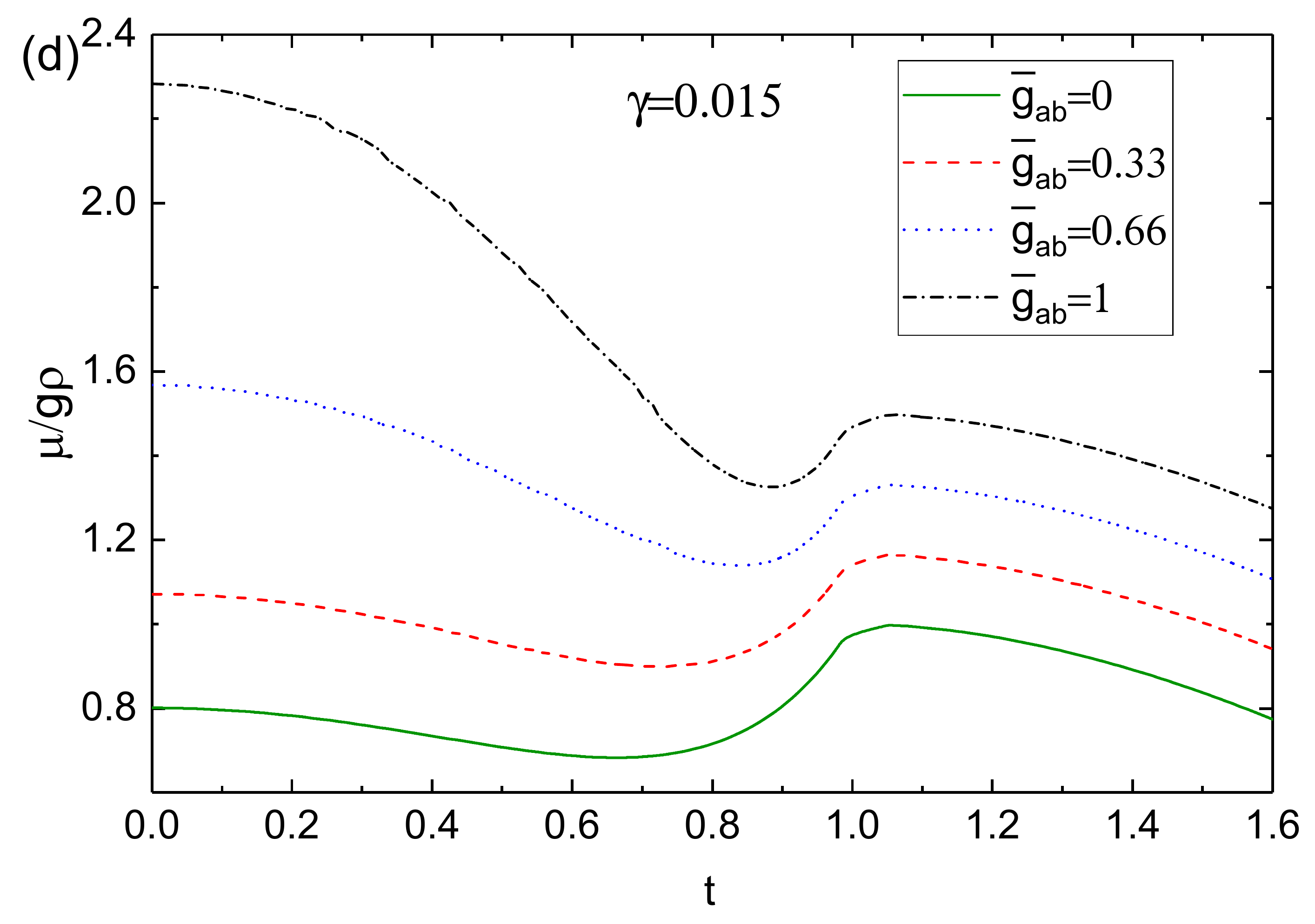}  }
\\
\end{minipage}
\hfill
\caption
{ Reduced chemical potential, $\mu/g\rho$, in the whole range of the dimensionless
temperatures $t=T/T_c$ for different values of $\overline{g}_{ab}$ and $\gamma$.
}
  \label{Fig5}
\end{figure}

In Fig. 5 we present the chemical potential on the whole range of temperatures $(t=T/T_c)$
for different values of $\overline{g}_{ab}$. It is seen that the modification of $\mu$ due to
$\overline{g}_{ab}$ is rather large both in Bose-condensed $(T<T_c)$ and normal $(T>T_c)$
phases.

\section{Discussion and conclusions}

We have developed a self-consistent mean-field theory for a binary homogeneous mixture of
two-component Bose systems. This theory, being conserving and gapless, imposes no restriction
on the gas parameter $\gamma$, and hence, it is valid for arbitrary strong interactions
$g_{ab}$. The theory satisfies the generalized HP theorem and takes into account anomalous
densities $\sigma_a$, $\sigma_b$, and $\sigma_{ab}$.
The presented approach is a kind of a self-consistent Hartree-Fock-Bogoliubov approximation,
hence it is the most general mean-field approximation. Therefore, as particular cases it 
includes other known mean-field approximations, such as the Shohno model, quadratic, and 
Bogoliubov approximations.
For numerical analysis, we have considered the balanced symmetric configuration of a
two-component mixture of Bose gases. We have obtained the phase diagram for this system at
zero as well as at finite temperatures for arbitrary gas parameters. The phase diagram at
zero temperature on the ($g_{ab}$, $\gamma$) plane shows that the system may remain stable
and miscible even at $g_{ab}/g_{aa}>1$, provided the anomalous densities are properly taken
into account. Comparing this phase diagram with that at finite temperature
($g_{ab}$, $\gamma$, $T$), we see that the finite temperature can transform the phase-
separated two-component BECs at $T=0$ to a miscible state. This conclusion is in good
agreement with the works by Roy \textit{et al.} \cite{roy15}, Ota \textit{et al.} \cite{ota20}, and Shi \textit{et al.}
\cite{shi2000}. Our numerical results are also in  good agreement with experimental works
\cite{kim20,Lee20}, although new experimental measurements for larger values of the
interspecies coupling and $\gamma$ are required.

The increase of the region of miscibility, due to the proper taking account of anomalous 
averages and temperature, can be understood remembering that these characteristics take 
into consideration the existence of quantum and temperature fluctuations. In the system, 
there are different competing factors. From one side, the repulsive interspecies 
interactions intend to separate the mixture. From the other side, larger interactions
induce larger anomalous averages, and the larger amount of uncondensed particles, which 
characterizes the increasing quantum fluctuations. Both quantum as well as thermal 
fluctuations are favorable for mixing. Under the given interactions, fluctuations 
facilitate the process of mixing. This is why the system may be immiscible, when 
fluctuations are absent but becomes miscible in the presence of fluctuations. As an 
example of the influence of thermal fluctuations, it is possible to consider the role 
of temperature in the thermodynamic miscibility conditions at weak interactions, when 
the free energies of mixed and separated states are compared. Taking into account that 
the difference in the entropy between the mixed and separated states is due to the mixing 
entropy, one obtains \cite{Yukalov_LP} the miscibility condition
$$
g_{ab} \; - \; \sqrt{g_{aa} \; g_{bb} } \; < \;
- \frac{2T}{\rho} \; \sum_i n_i \; \ln n_i \qquad
\left( n_i \equiv \frac{N_i}{N} \right) \; ,             
$$
where $\rho$ is the average density of the mixed system. As is evident, finite 
temperatures do facilitate the mixing, so that at zero temperature the system can be 
immiscible, while at finite temperature it can become miscible.     

The definition of the energy dispersion as well as the identification of points of instability,
introduced in Secs. II and III, requires some clarification. Actually, for this purpose we have
exploited first-order Green's functions \re{greeninv}. However, strictly speaking, these
parameters should be related to the poles of the full interacting Green's function, given by
the Dyson equation $\widehat{G}^{-1}=\widehat{G}^{-1}_{0}-\widehat{\Sigma}$, where
$\widehat{G}_{0}$ is the  ``noninteracting" Green's function and $\widehat{\Sigma}$ is the self-energy
operator \ci{stoof}. In the present approach $\widehat{G}_{0}$ does not coincide with that of
an ideal gas, but effectively takes into account two-body interactions via variational
parameters $[X_1, \dots, X_6]$.

It will be quite interesting to study a nonsymmetric, e.g.,  imbalanced two-component Bose mixture,
where quasimagnetic transitions may also take place \cite{ota20}. Moreover, as has been
recently claimed by Naidon and Petrov \cite{Petrov20}, for unequal interspecies interaction
or unequal masses the mixed phase can form bubbles with a tunable population. This will be
the next task for the application of our theory, since the ground-state physics can be
qualitatively understood from the arguments valid for a homogeneous system. However, in the case
of real systems in a trap phase separation can be suppressed in the inhomogeneous system due
to quantum pressure effects \ci{Patisson}.

\section*{Acknowledgments}

We are indebted to B. Tanatar for useful discussions. This work is supported by the Ministry
of Innovative Development of the Republic of Uzbekistan and the Scientific and Technological
Research Council of Turkey (TUBITAK) under Grant No. 119N689. A.R. acknowledges partial support
from the Academy of Sciences of the Republic of Uzbekistan.

\clearpage
\appendix
\section{The Green's functions and the densities}
\label{sec:A}
Here we present the explicit expressions for the Green's function
$G_{ij}(\omega_n,\vec{k})$ ($i,j=1- 4$) and the related densities.
By inversion of $G^{-1}_{ij}(\omega_n,\vec{k})$ given in Eq. \re{greeninv} one obtains
\bea
&& G_{11}(\omega_n,\vec{k})=\frac{W_2W_3W_4+W_2\omega_n^2-X_6^2W_3}{{\tilde D}}, \\
&& G_{12}(\omega_n,\vec{k})=-\frac{(W_3W_4+\omega_n^2+X_5X_6)\omega_n}{{\tilde D}}, \\
&& G_{13}(\omega_n,\vec{k})=-\frac{W_2X_5 W_4-X_6^2X_5-\omega_n^2 X_6}{{\tilde D}}, \\
&& G_{14}(\omega_n,\vec{k})=-\frac{(X_5W_2+W_3X_6)\omega_n}{{\tilde D}}, \\
&& G_{22}(\omega_n,\vec{k})=\frac{W_1W_3W_4+W_1\omega_n^2-W_4X_5^2}{{\tilde D}},
\eea
\bea
&& G_{23}(\omega_n,\vec{k})=-\frac{(X_6 W_1+X_5 W_4)\omega_n}{{\tilde D}}, \\
&& G_{24}(\omega_n,\vec{k})=-\frac{W_3X_6 W_1-X_6 X_5^2-\omega_n^2X_5}{{\tilde D}}, \\
&& G_{33}(\omega_n,\vec{k})=\frac{W_4W_1W_2+W_4\omega^2-X_6^2W_1}{{\tilde D}},\\
&& G_{34}(\omega_n,\vec{k})=-\frac{(W_1W_2+\omega_n^2+X_5 X_6)\omega_n}{{\tilde D}}, \\
&& G_{44}(\omega_n,\vec{k})=\frac{W_3W_1W_2+W_3\omega_n^2-X_5^2 W_2}{{\tilde D}},
\eea
where ${\tilde D}=(\omega_n^2+\omega_1^2)(\omega_n^2+\omega_2^2)$, $\omega_n=2\pi n T$,
$\omega_{1,2}$ are given in the main text, and $ W_1=\veps_a(k)+X_1$, $ W_2=\veps_a(k)+X_2$,
$W_3=\veps_b(k)+X_3$, $W_4=\veps_b(k)+X_4$, $X_5=X_{13}$, and  $X_6=X_{24}$.

Below we present explicit expressions for the densities defined in
Eqs. \re{rho1a.rho1b}--\re{RhoSigab}:
\begin{widetext}
\bea
\rho_{1a} && =\frac{1}{2\sqrt{D}\omega_1(k)\omega_2(k)V}\sum_k\left( [X_6^2W_3+W_2\omega_1^2(k)-W_2W_3W_4-W_1W_3W_4\right.+X_5^2W_4 +\nonumber \\
&&+W_1\omega_1^2(k)]\omega_2(k){\tilde W}_1(k)+[-X_6^2W_3-W_2\omega_2^2(k)+W_2W_3W_4+ \nonumber \\
&& \left. +W_1W_3W_4-X_5^2W_4-W_1\omega_2^2(k)]\omega_1(k){\tilde W}_2(k) \right),
\label{App_rho}
\eea
\bea
\rho_{1b} && =\frac{1}{2\sqrt{D}\omega_1(k)\omega_2(k)V}\sum_k\left( [X_6^2W_1+W_4\omega_1^2(k)-W_1W_2W_3-W_1W_2W_4+\right.X_5^2W_2+ \nonumber \\
&&+W_3\omega_1^2(k)]\omega_2(k){\tilde W}_1(k)+[-X_6^2W_1-W_4\omega_2^2(k)+W_1W_2W_3+ \nonumber \\
&&+\left. W_1W_2W_4-X_5^2W_2-W_3\omega_2^2(k)]\omega_1(k){\tilde W}_2(k) \right), 
\eea
\bea
\sigma_a && =\frac{1}{2\sqrt{D}\omega_1(k)\omega_2(k)V}\sum_k\left( [X_6^2W_3+W_2\omega_1^2(k)-W_2W_3W_4+W_1W_3W_4-\right. \nonumber \\
&& -X_5^2W_4-W_1\omega_1^2(k)]\omega_2(k){\tilde W}_1(k)+[-X_6^2W_3-W_2\omega_2^2(k)-W_1W_3W_4- \nonumber \\
&& -\left.W_2W_3W_4-W_2\omega_2^2(k)+X_5^2W_4+W_1\omega_2^2(k)]\omega_1(k){\tilde W}_2(k) \right), 
\eea
\bea
\sigma_b && =\frac{1}{2\sqrt{D}\omega_1(k)\omega_2(k)V}\sum_k\left( [X_6^2W_1+W_4\omega_1^2(k)+W_1W_2W_3-W_1W_2W_4-\right. \nonumber \\
&& -X_5^2W_2-W_3\omega_1^2(k)]\omega_2(k){\tilde W}_1(k)+
[-X_6^2W_1-W_4\omega_2^2(k)-W_1W_2W_3+ \nonumber \\
&& \left. +W_1W_2W_4+X_5^2W_2+W_3\omega_2^2(k)]\omega_1(k){\tilde W}_2(k) \right), 
\eea
\bea
\rho_{ab} && =\frac{1}{\sqrt{D}\omega_1(k)\omega_2(k)V}\sum_k\left( [-X_6^2X_5+W_1W_3X_6+W_2W_4X_5+\omega_1^2(k)X_6+\right.
 \nonumber \\
&& + X_5\omega^2_1(k)-X_5^2X_6]\omega_2(k){\tilde W}_1(k)+[X_6^2X_5-W_1W_3X_6-W_2W_4X_5- \nonumber \\
&& -\left.\omega_2^2(k)X_6-X_5\omega^2_2(k)+X_5^2X_6]\omega_1(k){\tilde W}_2(k) \right),
 \eea
 \bea
\sigma_{ab} && =\frac{1}{\sqrt{D}\omega_1(k)\omega_2(k)V}\sum_k\left( [-X_6^2X_5-W_1W_3X_6+W_2W_4X_5+\omega_1^2(k)X_6-\right. \nonumber \\
&& - X_5\omega^2_1(k)+X_5^2X_6]\omega_2(k){\tilde W}_1(k)+ [X_6^2X_5+W_1W_3X_6-W_2W_4X_5-\nonumber \\
&&-\left.x\omega_2^2(k)X_6-X_5\omega^2_2(k)+X_5^2X_6]\omega_1(k){\tilde W}_2(k) \right), 
\label{App_sig}
 \eea
 \end{widetext}
where  ${\tilde W}_{1,2}(k)={1}/{2}+f(\omega_{1,2}(k))$, and $D$, $\omega_{1,2}(k)$ are
given in Eqs. \re{dispX5X6} -- \re{Eab}. 

Note that, in practical calculation of momentum integrals, adequate counter-terms should be
included similar to $c_\rho=-1/2$  (for $\rho_{1}$), and $c_\sigma=\Delta/2\veps_k$ 
(for $\sigma$)  used in the  one-component case.

\newpage

\bb{99}

\bi{myat97}  C. J. Myatt, E. A. Burt, R. W. Ghrist, E. A. Cornell, and C. E. Wieman,
Phys. Rev. Lett. \textbf{78}, 586, (1997).

\bi{hall98} D. S. Hall, M. R. Matthews, J. R. Ensher, C. E. Wieman, and E. A. Cornell,
Phys. Rev. Lett. \textbf{81}, 1539 (1998).

\bi{Petrov15} D. S. Petrov,
Phys. Rev. Lett. \textbf{115}, 155302 (2015).

\bi{Timmerman} E. Timmermans,
Phys. Rev. Lett. \textbf{81}, 5718 (1998).

\bi{papp2008} S. B. Papp, J. M. Pino, R. J. Wild, S. Ronen, C. E. Weiman, D. S. Jin and E. A. Cornell,
Phys. Rev. Lett. \textbf{101}, 135301 (2008).

\bi{papp20082} S. B. Papp, J. M. Pino, and C. E. Wieman,
Phys. Rev. Lett. \textbf{101}, 040402 (2008).

\bi{burchanti202} A. Burchianti, C. D'Errico, S. Rosi, A. Simoni, M. Modugno, C. Fort,
and F. Minardi,
Phys. Rev. A \textbf{98}, 063616 (2018).

\bi{kim20} J. H. Kim, D. Hong, and Y. Shin,
Phys. Rev. A \textbf{101}, 061601(R) (2020).

\bi{Lee20} K. L. Lee, N. B. Jørgensen, L. J. Wacker, M. G. Skou, K. T. Skalmstang,
J. J. Arlt, and N. P. Proukakis,
New J. Phys. \textbf{20}, 053004 (2018).

\bi{wacher15} L. Wacker, N. B. Jørgensen, D. Birkmose, R. Horchani, W. Ertmer, C. Klempt,
N. Winter, J. Sherson, and J. J. Arlt,
Phys. Rev. A \textbf{92}, 053602 (2015).

\bi{wang15} F. Wang, X. Li, D. Xiong, and D. Wang,
J. Phys. B: At. Mol. Opt. Phys. \textbf{49}, 015302 (2016).

\bi{cabrera18} C. R. Cabrera, L. Tanzi, J. Sanz, B. Naylor, P. Thomas, P. Cheiney, and
L. Tarruell,
Science \textbf{359}, 301 (2018).

\bi{fava18} E. Fava, T. Bienaimé, C. Mordini, G. Colzi, C. Qu, S. Stringari, G. Lamporesi,
and G. Ferrari,
Phys. Rev. Lett. \textbf{120}, 170401 (2018).

\bi{ozaki10} T. Ozaki, and T. Nikuni,
J. Phys. Soc. Jpn. \textbf{81}, 024001 (2010).

\bi{shi2000} H. Shi, W. M. Zheng, and S. T. Chui,
Phys. Rev. A \textbf{61}, 063613 (2000).

\bi{schayback13} B. V. Schaeybroeck,
Physica A \textbf{392}, 3806 (2013).

\bi{roy15} A. Roy and D. Angom,
Phys. Rev. A \textbf{92}, 011601(R) (2015).

\bi{sasaki09} K. Sasaki, N. Suzuki, D. Akamatsu, and H. Saito,
Phys. Rev. A \textbf{80}, 063611 (2009).

\bi{takeuchi13} H. Takeuchi, N. Suzuki, K. Kasamatsu, H. Saito, and M. Tsubota,
Phys. Rev. A \textbf{81}, 094517 (2010).

\bi{tickhor13} C. Ticknor,
Phys. Rev. A \textbf{88}, 013623 (2013).

\bi{kim02} J. G. Kim and E. K. Lee,
Phys. Rev. E \textbf{65}, 066201, (2002).

\bi{hryhorchak} O. Hryhorchak and V. Pastukhov,
J. Low Temp. Phys. \textbf{202}, 219 (2021).

\bi{ota20} M. Ota and S. Giorgini,
Phys. Rev. A \textbf{102}, 063303 (2020).

\bi{boudjuma} A. Boudjemâa,
Phys. Rev. A \textbf{ 97}, 033627 (2018).

\bi{andersen} J. O. Andersen,
Rev. Mod. Phys. \textbf{76}, 599 (2004).

\bi{pines1959} N. M. Hugenholtz and D. Pines,
Phys. Rev. \textbf{116}, 489 (1959).

\bi{watabe} S. Watabe,
Phys. Rev. A \textbf{103}, 053307 (2021).

\bi{Yukalov_PRE} V. I. Yukalov,
Phys. Rev. E \textbf{72}, 066119 (2005).

\bi{Yukalov_PLA} V. I. Yukalov,
Phys. Lett. A \textbf{359}, 712 (2006).

\bi{yukalovannals} V. I. Yukalov,
Ann. Phys. (N.Y.) \textbf{323}, 461 (2008).

\bi{ourAniz1} A. Khudoyberdiev, A. Rakhimov, and A. Schilling,
New J. Phys. \textbf{19}, 113002 (2017).

\bi{ourannalshJs} A. Rakhimov, S. Mardonov, and E. Ya. Sherman,
Ann. Phys. (N.Y.) \textbf{326}, 2499 (2011).

\bi{mysolo} A. Rakhimov,
Phys. Rev. A \textbf{102}, 063306 (2020).

\bi{ourctan2} A. Rakhimov, T. Abdurakhmonov and B. Tanatar,
J. Phys.: Condens. Matter \textbf{33}, 465401 (2021).

\bi{ourAniz2part1} A. Rakhimov, A. Khudoyberdiev, L. Rani, and B. Tanatar,
Ann. Phys. (N.Y.) \textbf{424}, 168361 (2021).

\bi{ourpra77} 
A. Rakhimov, Chul Koo Kim, Sang - Hoon Kim and Jae Hyung Yee,
Phys. Rev. A    \textbf{77}, 033626  (2008).

\bi{ouraniz2part2} A. Rakhimov, A. Khudoyberdiev, and B. Tanatar,
Int. J. Mod. Phys. B \textbf{35}, 2150223 (2021).

\bi{navon2011} N. Navon, S. Piatecki, G. Gunter, B. Rem. T. Nguyen,
Phys. Rev. Lett. \textbf{107}, 135301 (2011).

\bi{lopes2017}R. Lopes, C. Eigen, A. Barker, K. Veibahn, M. Vincent, 
Phys. Rev. Lett. \textbf{118}, 210401 (2017).

\bi{Nepomnyash}
Y. A. Nepomnyashchii,
J. Exp. Theor. Phys. {\textbf 43}, 559 (1976).

\bi{pinto} M. B. Pinto, R. O. Ramos, and F. F. de Souza Cruz,
Phys. Rev. A \textbf{74}, 033618 (2006).

\bi{stev81} P. M. Stevenson,
Phys. Rev. D \textbf{23}, 2916 (1981).

\bi{ouryee04} A. Rakhimov and J. H. Yee,
Int. J. Mod. Phys. A \textbf{19}, 1589 (2004).

\bi{Yukalov_PPN} V. I. Yukalov,
Phys. Part. Nucl. \textbf{50}, 141 (1019).

\bi{yukphys21} V. I. Yukalov and E. P. Yukalova,
Physics \textbf{3}, 829 ( 2021).

\bi{yukechaya} V. I. Yukalov,
Phys. Part. Nucl. \textbf{42}, 460 (2011).

\bi{HohenbergM} P. C. Hohenberg and P. C. Martin,
Ann. Phys.  (N.Y.) \textbf{34}, 291 (1965).

\bi{stevenson} P. M. Stevenson,
Phys. Rev. D \textbf{32}, 1389, (1985).

\bi{orkl1} H. Kleinert, Z. Narzikulov, and A. Rakhimov,
Phys. Rev. A \textbf{85}, 063602 (2012).

\bi{ourkl2} H. Kleinert, Z Narzikulov, and A. Rakhimov,
J. Stat. Mech. P01003 (2014).

\bi{Yukalov_LP} V. I. Yukalov,
Laser Phys. \textbf{26}, 062001 (2016).

\bi{stev42} I. Stancu and P. M. Stevenson, 
Phys. Rev. D \textbf{42}, 2710, (1989).

\bi{etawenfa} L. Wen, W. M. Liu, Y. Y. Cai, J. M. Zhang, and J. P. Hu,
Phys. Rev. A \textbf{ 85}, 043602 (2012).

\bi{dipolarbrazileta} R. K. Kumar, P. Muruganandam, L. Tomio, and A. Gammal,
J. Phys. Commun. \textbf{1}, 035012 (2017).

\bi{Shohno} N. Shohno, 
Prog. Theor. Phys. \textbf{31}, 553 (1964).

\bi{Gavoret}
G. Gavoret,
Ann. Phys. (N.Y.) {\bf 28}, 349 (1964).

\bi{Yukalov_APPA}
V. I. Yukalov,
Acta Phys. Pol. A {\bf 57}, 295 (1980).

\bi{larsen1963} D. M. Larsen,
Ann. Phys.  (N.Y.) \textbf{24}, 89 (1963).

\bi{LHY} T. D. Lee, K. Huang, and C. N. Yang
Phys. Rev. \textbf{136}, 1135 (1957).

\bi{stoof} H. T. C. Stoof, K. B. Gubbels, and D. B. M. Dickerscheid,
\textit{Ultracold Quantum Fields} (Springer, Berlin, 2008).

\bi{Petrov20} P. Naidon and D. S. Petrov,
Phys. Rev. Lett. \textbf{126}, 115301 (2021).

\bi{Patisson} R. W. Pattinson, T. P. Billam, S. A. Gardiner, D. J. McCarron, H. W. Cho,
S.L. Cornish, N. G. Parker, and N. P. Proukakis,
Phys. Rev. A \textbf{87}, 013625 (2013).

\eb

\edc